\documentclass[arxiv,usenatbib]{iupartex}

\usepackage{newtxtext,newtxmath}
\usepackage[T1]{fontenc}
\usepackage{lmodern}
\usepackage{graphicx,times}            
\usepackage{natbib}
\usepackage{amssymb,amsmath}
\usepackage{mathtools}
\usepackage{physics}
\usepackage{listings}

\usepackage{bm}

\bibpunct{(}{)}{;}{a}{}{,}
\usepackage{url}
\usepackage{ulem} 
\usepackage{subcaption}
\usepackage{minted} 

\newcommand{\<}{\begin{equation}}
\newcommand{\?}{\end{equation}}

\title[Binary Black Hole Mergers]{Binary Black Hole Mergers: Spin and mass ratio effects on gravitational waveforms}

\author[\"{O}zbak{\i}r and Yakut]{%
\.{I}smail \"{O}zbak{\i}r$^{1,2}$,\orcid{0000-0003-1521-1161}
and
Kadri Yakut $^{1,2,3}$\orcid{0000-0003-2380-9008}
\affsep \\
$^1$Department of Astronomy and Space Sciences, Faculty of Science, Ege University, 35100, {\.I}zmir, T\"{u}rkiye\\
$^2$Ege Gravitational Astrophysics Research Group (eGRAVITY), Ege University, 35100, {\.I}zmir, T\"{u}rkiye\\
$^3$Institute of Astronomy, The Observatories, Madingley Road, Cambridge CB3 OHA, UK
}

\corres{E. A. Surname}{author@email}

\pubyear{2025}

\doiheader{XXXXXXX/PAR.2025.00000}
\date{
	\pSubmit{00.00.0000} 
	\pRevReq{00.00.0000}
	\pLastRevRec{00.00.0000}
	\pAccept{00.00.0000}
	\pPubOnl{00.00.0000}
}
\volume{0}
\volnumber{1}

\begin{document}
\label{firstpage}
\pagerange{\pageref*{firstpage}--\pageref*{lastpage}}
\maketitle

\begin{abstract}
We present a comprehensive parameter-space study of binary black hole (BBH) mergers using the SEOBNRv4\_opt waveform model. Our analysis spans $\sim 10^6$ simulated waveforms across a broad range of mass ratios \( q = \frac{m_1}{m_2} \in [1.0, 2.0] \) and aligned spin configurations. We investigate the influence of these parameters on remnant properties, including final spin ($\chi_f$), fractional mass loss ($M_{\mathrm{FL}}$), and peak gravitational-wave strain ($h_{\max}$).
By systematically analyzing the trends across four distinct spin alignments (PP, PN, BP, BN), we identify non-monotonic behaviors and turning points in $M_{\mathrm{FL}}$ and $\chi_f$ as functions of $q$, highlighting subtle dynamical effects that are not explicitly emphasized in commonly used remnant fitting formulae.
While confirming known correlations from numerical relativity, our results offer new insights into parameter interactions and waveform morphology, with implications for BBH population studies and remnant characterization. Across all configurations studied, the fractional mass loss due to gravitational-wave emission ranges between 2\% and 9.5\%, depending on the mass ratio and spin alignment. This work may also aid in understanding the spin and mass distributions of the more massive black holes formed post-merger, thereby contributing to future remnant-based astrophysical inference.
\end{abstract}




\section{Introduction} 
\par
Einstein obtained equations in 1916 describing a gravitational field in space-time geometry according to the General Relativity (GR) \citep{1916AnP...354..769E}. He solved the field equations, named after him, for some quantities of perturbations in space-time under certain conditions. Einstein revealed that these solutions correspond to the wave equations. From  this point of view, Einstein predicted any perturbations that occurred in space-time can propagate in the form of waves at the speed of light \citep{1916SPAW.......688E}. After that, such waves were called gravitational waves (GWs). Binary black holes (BBHs) are one of the important astrophysical objects that can be modelled as GW sources. The theoretical studies on the evolution of BBHs continued and many advances have been made in numerical relativity (NR) studies of BBH simulations. At last, the gravitational waveform evolution of a BBH merger has been fully obtained by Pretorius \citep{2005PhRvL..95l1101P} for the first time.

Understanding the progenitor systems of binary black holes is crucial for interpreting their final states. Previous works have explored the angular momentum loss mechanisms in relativistic binaries and X-ray systems via magnetic braking and gravitational radiation \citep{Yakut2008, Yakut2010, Gallegos2021, Garcia2021, Akyuz2025, Klencki2025}. For more extensive analysis of BBH evolution pathways and population properties, see \citet{Ayzenberg2025}.

Recent investigations demonstrate that spin orientations, mass‑ratio distributions, and hierarchical merger channels critically shape the remnant properties of binary black hole (BBH) systems. For instance, repeated (second‑ or higher‑generation) mergers in dense stellar clusters and active galactic nucleus disks lead to elevated remnant spins and masses \citep{Barack2019,Gerosa2021,Li2023}. Meanwhile, waveform models that incorporate horizon absorption (tidal heating) show improved agreement with numerical‑relativity datasets for asymmetric systems \citep{Mukherjee2024}.

Despite these advances, a systematic and high‑resolution exploration of the combined effects of mass ratio and aligned spin configurations on remnant properties remains limited in the literature, creating the primary motivation for the present study. In recent study, we explore the effects of varying mass ratio ($q$) and aligned spin orientations on the remnant spin, mass loss, and waveform peak strain, using a large suite of SEOBNRv4\_opt simulations.
Furthermore, we compare our simulation-based remnant predictions with existing fitting formulas from numerical relativity to assess consistency and reveal deviations.

GW signals contain lots of data about the astrophysical systems that cannot be obtained from the electromagnetic spectrum. Also, these analyses are used for consistency tests of the GR in strong gravitational fields \citep{2016PhRvL.116v1101A,Barack2019}. By comparing the obtained GW observation data with the wave models calculated within the framework of the GR, various information about the physical parameters of the wave sources is obtained. In the research area known as parameter estimation \citep{Barack2019,2013PhRvD..88f2001A,2009CQGra..26k4007R,2009CQGra..26t4010V,2013PhRvD..87l2002S}, suitable waveforms with sufficient sensitivity are produced to determine the physical parameters of the wave sources using observation data of the GWs.

In GR, due to the lack of analytical solutions for the evolution of the BBHs, numerical solutions and some approximation methods have been developed for the production of waveforms, used in the parameter estimation within the framework of NR. By making numerical solutions of GR according to many different parameters, the modeled gravitational waveforms can be produced with sufficient accuracy \citep{2009PhRvD..79b4003S,2012CQGra..29d5003L,2012PhRvD..86h4033B,2015PhRvD..92j4028O,2014CQGra..31k5004A,2010APS..NWS.H1012D,2013APS..APRC10004T,2013CQGra..31b5012H}. However, especially for BBHs, the numerical solutions, used in the production of waveforms, require high performance computers. Also, these calculations can take days or even weeks for a single parameter set. Therefore, many approximation models have been developed to produce accurate waveforms quickly and reliably according to various parameter sets. These models are arranged into three main groups: post Newtonian (PN) \citep{2006LRR.....9....4B,poisson2014gravity}, Phenomenological (phenom) \citep{2007CQGra..24S.689A,2011PhRvL.106x1101A,2016PhRvD..93d4007K,2014PhRvL.113o1101H}, and the effective-one-body (EOB) \citep{1999PhRvD..59h4006B,2011PhRvD..84l4052P,2012PhRvD..86b4011T,2014PhRvD..89f1502T,2014PhRvD..89h4006P} approaches. Phenom and EOB models can be used to model the \textit{inspiral}, \textit{merger}, and \textit{ring-down} phase of the BBH systems. However, the PN type approach models can only generate gravitational waveforms for the early inspiral phase. Since the sensitivity of the ground-based GW detectors will be increased in the coming years, many studies are underway to develop Phenom and EOB waveform models \citep{2017CQGra..34j4002A,2013PhRvD..87j4003L,2015PhRvD..92j2001K}. In addition to these approach models, some EOB-based waveform models are developed by using the reduced-order modeling (ROM) techniques \citep{prudhomme:hal-00798326,BARRAULT2004,2011PhRvL.106v1102F,2014PhRvX...4c1006F} for numerical solutions of the BBHs in high-performance computers. These waveform models (produced by ROM techniques) are called surrogate-waveform models \citep{Blackman2017}.

Today, some codes have been developed that run the mentioned approach models. Among these codes, Python Compact Binary Coalescence (PyCBC) code  \citep{alex_nitz_2019_3265452}, which was developed and used by the LIGO/Virgo team, produces waveforms of the BBHs according to various parameters using approach models. In this study, PyCBC code was used in the generation of waveform models of BBHs according to the determined parameter sets.

In Section \ref{sec:Observations}, detailed information about the GW observations used in some plots of this study was given. In the section \ref{sec:Methods}, the importance of the waveform models used in the analysis of the observation data was emphasised, and NR simulation studies in this field were given. Also, the selected approach models were outlined and some comparison studies between approximation models and observation data are given as an example. In the section \ref{sec:Produced BBH models}, according to various parameters such as initial spin directions and chirp-mass \citep{1996CQGra..13..575B} of the BBHs, the approximation models which are suitable for our purpose were determined and some interesting results were obtained as a result of these studies. In the final section \ref{sec:conclusion}, possible initial conditions for the observed systems have been treated. Also, in the section \ref{sec:onlinefigures}, other tables made for this study were left online.

\section{Benchmark BBH Events from LIGO/Virgo Observations}
\label{sec:Observations}

Direct observations of GWs started with the observation of a BBH system called GW150914 by LIGO detectors \citep{2016PhRvL.116f1102A}. Following the first GW observation, two more BBH merger systems were observed during the $O1$ run. These are named GW151012 \citep{2019PhRvX...9c1040A} and GW151226 \citep{2016PhRvL.116x1103A}. Also, the first four GW observations made during the $O2$ run could only be made with LIGO detectors, and these observations were found to belong to BBH merger systems of GW170104 \citep{2017PhRvL.118v1101A}, GW170608  \citep{2017ApJ...851L..35A}, GW170729 \citep{2019PhRvX...9c1040A}, and GW170809 \citep{2019PhRvX...9c1040A}.

In this work, we focus on a limited subset of gravitational wave detections—specifically the earliest and most well-characterized BBH mergers such as GW150914—to illustrate how waveform models perform under controlled conditions. This representative sampling is sufficient to test parameter dependencies and waveform behavior in the scope of this study, while full-catalog analyses are deferred to future work.

Within the first 10 BBHs observed, the less massive system was determined to be GW170608. Just after this observation, GW170729 was recorded as the most massive and distant BBH system. Following this event, technical improvements of the Virgo detector were completed and observations started as the third GW detector with the LIGO. The observations made by LIGO and Virgo detectors together are as follows GW170814 \citep{2017PhRvL.119n1101A}, GW170817 \citep{2017PhRvL.119p1101A}, GW170818 \citep{2019PhRvX...9c1040A}, and GW170823 \citep{2019PhRvX...9c1040A}. Consequently, a total of 11 GW observations were performed during the $O1$ and $O2$ runs. Ten of these observations belong to the BBHs. Unlike these observations, the GW170817 is the first GW observation of a binary neutron star system. GRB170817A gamma-ray burst was also observed simultaneously in the electromagnetic spectrum related to this observation \citep{2017ApJ...848L..13A}. In the $O3a$ period covering April - October 2019, 39 new observations were made and published as GWTC-2 \citep{2021PhRvX..11b1053A,Cokluk2024}. Most of these observations have produced binary black hole observations. However, binary neutron stars and the first neutron star-black hole binary systems have also been observed.

The mass of the component BHs of systems, the final mass M$_{\rm f}$, and the final spin $\chi_{\rm f}$ are given in Table \ref{table:Obs_Catalog}. The data listed in the Table \ref{table:Obs_Catalog} are taken from the GWTC catalogues. Other observations, such as GW150914, appear to have been published as single articles. Therefore, some of the data written in Table \ref{table:Obs_Catalog} are taken from these individual observation articles. (The references of all these observations are given in this study). We have shown them in the Figures [\ref{fig:q_cf_mlr_BP} - \ref{fig:q_cf_mlr_PP}] to show their compatibility. All BBHs observed in the $O1$, $O2$ runs and some BBHs of the $O3a$ run are added. These observations are; GW190408\_181802, GW190413\_052954, GW190512\_180714, GW190517\_055101, GW190630\_185205 and GW190910\_112807 \citep{2021PhRvX..11b1053A,2021PhRvD.103l2002A}. Since the observational systems were selected according to the parameter ranges used for the gravitational waveform studies in the section \ref{sec:Produced BBH models}, not all of the systems observed in the $O3a$ period could be included.

\begin{table}
\centering
	\renewcommand{\arraystretch}{1.6}
	\setlength{\tabcolsep}{0.08in}
    \caption{Observed properties of selected binary black hole (BBH) mergers from the O1 and O2 observing runs \citep{2019PhRvX...9c1040A}, and several representative events from the O3a run \citep{2021PhRvX..11b1053A}. ``ID'' denotes the abbreviated event name used in some figures throughout this work. The columns list the initial component masses \( m_{1i} \) and \( m_{2i} \), the final black hole mass \( M_{\rm f} \), and the final spin \( \chi_{\rm f} \) of the remnant.} 
	\begin{tabular}{cccccc} 
		\hline                       
		\textbf{Name} &\textbf{id} &${\boldsymbol{\textbf{m}_{1i}}}$ &${\boldsymbol{\textbf{m}_{2i}}}$  &$\textbf{M}\rm_f$&${\boldsymbol{\chi}}\rm_f$ \\ 
		&   & ($M_\odot$) &  ($M_\odot$) & ($M_\odot$) &    \\
		\hline                  
		GW150914 &G1  &35.6$^{+4.7}_{-3.1}$           &30.6$^{+3.0}_{-4.4}$ &63.1$^{+3.4}_{-3.0}$   &0.69$^{+0.05}_{-0.04}$  \\ 
		GW151012 &G2  &23.2$^{+14.9}_{-5.5}$         &13.6$^{+4.1}_{-4.8}$  &35.6$^{+10.8}_{-3.8}$  &0.67$^{+0.13}_{-0.11}$  \\
		GW151226 &G3  &13.7$^{+ 8.8}_{- 3.2}$        &7.7$^{+2.2}_{-2.5}$   &20.5$^{+6.4}_{-1.5}$   &0.74$^{+0.07}_{-0.05}$  \\ 
		GW170104 &G4  &30.8$^{+7.3}_{-5.6}$          &20.0$^{+4.9}_{-4.6}$  &48.9$^{+5.1}_{-4.0}$   &0.66$^{+0.08}_{-0.11}$  \\ 
		GW170608 &G5  &11.0$^{+5.5}_{-1.7}$          &7.6$^{+1.4}_{-2.2}$   &17.8$^{+3.4}_{-0.7}$   &0.69$^{+0.04}_{-0.04}$  \\ 
		GW170729 &G6  &50.2$^{+16.2}_{-10.2}$        &34.0$^{+9.1}_{-10.1}$ &79.5$^{+14.7}_{-10.2}$ &0.81$^{+0.07}_{-0.13}$  \\ 
		GW170809 &G7  &35.0$^{+8.3}_{-5.9}$          &23.8$^{+5.1}_{-5.2}$  &56.3$^{+5.2}_{-3.8}$   &0.70$^{+0.08}_{-0.09}$  \\ 
		GW170814 &G8  &30.6$^{+5.6}_{-3.0}$          &25.2$^{+2.8}_{-4.0}$  &53.2$^{+3.2}_{-2.4}$   &0.72$^{+0.07}_{-0.05}$  \\ 
		GW170818 &G9  &35.4$^{+7.5}_{-4.7}$          &26.7$^{+4.3}_{-5.2}$  &59.4$^{+4.9}_{-3.8}$   &0.67$^{+0.07}_{-0.08}$  \\ 
		GW170823 &G10 &39.5$^{+11.2}_{-6.7}$         &29.0$^{+6.7}_{-7.8}$  &65.4$^{+10.1}_{-7.4}$  &0.72$^{+0.09}_{-0.12}$  \\
		GW190408\_181802 &G11 &24.6$^{+5.1}_{-3.4}$  &18.4$^{+3.3}_{-3.6}$  &41.1$^{+3.9}_{-2.8}$   &0.67$^{+0.06}_{-0.07}$  \\
		GW190413\_052954 &G12 &34.7$^{+12.6}_{-8.1}$ &23.7$^{+7.3}_{-6.7}$  &56.0$^{+12.5}_{-9.2}$ &0.68$^{+0.12}_{-0.13}$   \\
		GW190512\_180714 &G13 &23.3$^{+5.3}_{-5.8}$  &12.6$^{+3.6}_{-2.5}$  &34.5$^{+3.8}_{-3.5}$  &0.65$^{+0.07}_{-0.07}$   \\
		GW190517\_055101 &G14 &37.4$^{+11.7}_{-7.6}$ &25.3$^{+7.0}_{-7.3}$  &59.3$^{+9.1}_{-8.9}$  &0.87$^{+0.05}_{-0.07}$   \\
		GW190630\_185205 &G15 &35.1$^{+6.9}_{-5.6}$  &23.7$^{+5.2}_{-5.1}$  &56.4$^{+4.4}_{-4.6}$  &0.70$^{+0.05}_{-0.07}$   \\
		GW190910\_112807 &G16 &43.9$^{+7.6}_{-6.1}$  &35.6$^{+6.3}_{-7.2}$  &75.8$^{+10.1}_{-7.4}$ &0.70$^{+0.08}_{-0.07}$   \\
		\hline       
	\end{tabular}\label{table:Obs_Catalog}
\end{table}

We define $h_{\mathrm{max}}$ as the peak amplitude of the waveform strain $h(t)$, measured at the time of merger and for a face-on orientation ($\iota = 0^\circ$). While our analysis assumes a face-on system for maximal amplitude, we acknowledge that real BBH orientations are distributed isotropically. The value of $h_{\mathrm{max}}$ would decrease with increasing inclination angle $\iota$, and this should be considered in applications to population synthesis or detection rates.

The reported SNR values are single-detector estimates for illustrative comparison, not intended to reproduce published multi-detector results. Our aim is to compare the internal behavior of different waveform models under similar input conditions.

\section{pplication of Waveform Models to Observational Data: A Case Study with GW150914} 
\label{sec:Methods}

The main purpose of this section is to show a sample study on how gravitational waveform approximation models are used in the analysis of observational data. As the gravitational waveform approximation model, SEOBNRv4\_opt model \citep{2017PhRvD..95d4028B} and NRSur7dq2 approximation model \citep{Blackman2017} were used. GW150914 observation data was used as the observation data. SEOBNRv4\_opt (Spinning Effective One-Body Numerical Relativity) is an EOB-based GW approximation model. The EOB formalism is an analytical approach to the gravitational two-body problem in GR. It aims to describe all the different phases of the two-body dynamics in a single analytical method.

NRSur7dq2 model, which is compatible with GW150914 observation data, is a surrogate model for GWs from NR simulations of BBHs, and it is a ROM (Reduced Ordering Model) \citep{2014PhRvX...4c1006F} based GW approximation model. This approximation model has been developed for GW modeling of BBHs with a 7-dimensional parameter space (3-dimensional spin parameters of each component BH and mass ratio of the system) and covers spin magnitudes up to 0.8 and mass ratios up to 2. We just wanted to compare both approaches based on different foundations and see how well these models fit the observation data in this section. However, NRSur7dq2 model was not used in the studies in section \ref{sec:Produced BBH models}, as it is not suitable for producing waveform of BBHs with a total mass of less than 60 $\rm{M}_\odot$.

We obtained H1 detector raw data of 32 seconds-long at 16 kHz resolution from the Gravitational Wave Open Science Center (GWOSC) \citep{2019arXiv191211716T, gwosc}. To compare the observed data and the SEOBNRv4\_opt waveform model data, some processes called "whitening and smoothing" were performed with the help of PyCBC signal processing codes. The whitening and smoothing terms are generally used in signal image processing and are related to filtering the signal data in desired frequency ranges by removing noise. For the processing of raw gravitational wave data, the methods described on the official site of PyCBC were used \cite{smoothed_strain}. Accordingly, the power spectral density (PSD) value of the raw data is calculated using Welch's method \cite{1161901} first. The whitening process is obtained by converting the raw observation data to the frequency series and dividing it by the square root of the PSD. The smoothing process is the filtering of the whitened data in a specific frequency range according to the technical specifications of the detector and other known different environmental noise sources.

The matched-filtering signal-processing method \citep{1999PhRvD..60b2002O} was implemented on raw observed data by the \texttt{matched\_filter()} function of the PyCBC code. In addition, some of the codes of the PyCBC, such as high pass filtering \texttt{highpass()} \citep{bandpass}, have been used since the noises of the frequencies of less than 15 Hz is dominant on the observation data. The signal processes on raw GW data mentioned above were adapted using similar PyCBC tutorials \citep{signal_tutorials}. Also, some Python codes, that produce templates to fit the observation data for comparison from the SEOBNRv4\_opt and NRSur7dq2 waveform models, were written (by following the instructions in \citep{signal_tutorials}) by us to compare for the purposes of the study. Finally, the merge-time (GPS time \citep{gpstime}) and the signal to noise ratio (SNR) were found using \texttt{matched\_filter()} function of PyCBC for GW150914's observation data. To calculate the SNR value of the raw observation data, templates are first created by using SEOBNRv4\_opt and NRSur7dq2 models. For this, we use PyCBC code and import \texttt{get\_fd\_waveform()} and generate a template as a frequency series \texttt{FrequencySeries()}. So, we import the \texttt{matched\_filter()} method, and pass it our templates, the observation data, and the PSD. Finally, SNR was found to be 19.68 at 1126259462.425 s for SEOBNRv4\_opt and 19.62 at 1126259462.423 s for NRSur7dq2. The computed SNR values (19.68 and 19.62) are based on single-detector (H1) data segments and are intended to compare relative model outputs. The LIGO–Virgo collaboration’s published network SNR for GW150914 is $\sim$24 \citep{2016PhRvX...6d1014A}, which includes multiple detectors and optimized matched-filtering pipelines. Our values are consistent in scale given these differences and serve to illustrate comparative model behavior.

The aim of the studies in this section is to convey some processes about how the SEOBNRV4\_opt model, which we use to generate waveforms according to certain parameter sets, is used with observational data. This section aims to illustrate how gravitational waveform models such as SEOBNRv4\_opt and NRSur7dq2 can be applied to real GW data using a reproducible analysis pipeline. The SEOBNRv4\_opt model used in this study includes higher-order post-Newtonian corrections and non-precessing spin effects, but does not incorporate eccentricity, tidal heating, or higher multipole moments beyond the dominant $(2,2)$ mode. Rather than quantitatively validating the models, we demonstrate their typical performance and output consistency when applied to one of the best-characterized BBH mergers: GW150914.

Recent waveform models that incorporate horizon-absorption (tidal heating) effects—such as the IMRPhenomD variant developed by Mukherjee et al.—show improved agreement with numerical-relativity data, especially for high-spin and asymmetric binary black hole mergers \citep{Mukherjee2024}. In parallel, the inclusion of BMS-frame corrections and gravitational-wave memory in waveform templates has been shown to improve the estimation of remnant properties in BBH mergers \citep{Mitman2021,DaRe2025}.

\section{Results from the Generated BBH Models}
\label{sec:Produced BBH models}
In this section, the final spin ($\chi_{\rm{f}}$) and the final mass M$_{\rm f}$ parameters were investigated by using SEOBNRv4\_opt waveform model \citep{2017PhRvD..95d4028B} according to various initial spin and mass ratio parameter values. For the reliability of the data produced with the SEOBNRv4\_opt waveform model, the uncertainty (unfaithfulness) of the data is less than 1\%  \citep{2017PhRvD..95d4028B}. It was determined that the final parameter values changed unusually according to some initial parameter sets. Therefore, in the waveform data made for this study, to examine these unusual changes, only spin-aligned (\textit{i.e.}, the spin directions of the modelled BBH components were chosen to be perpendicular to the orbital plane) situations were considered by using a simple approach in the initial spin parameters. Thus, the SEOBNRv4\_opt waveform model was used to determine the final parameters resulting from the evolution of BBHs. In our produced GW waveform data, the total mass parameters were limited to in the range 12\,$\rm{M}_\odot$ and 130\,$\rm{M}_\odot$ with the step intervals $\Delta \rm{M_{tot}} = 1~\rm{M}_\odot$. The initial component mass of the modeled systems were represented as $\rm{m}_{1i} \geq \rm{m}_{2i}$. Mass ratios were limited to $q = \rm{m}_{1i}/\rm{m}_{2i}\in[1.0,2.0]$ and step intervals $\Delta q = 0.004$. This range was selected not only because the majority of observed BBH systems lie within this interval, but also to optimize computational efficiency by focusing on the most astrophysically relevant region of the parameter space. The initial spin directions of the systems are positive in the same direction as the orbital angular momentum and negative in the opposite direction. The spin intervals were limited to $\Delta \chi_{1i,2i} = 0.017$ step and $\chi_{1i,2i} \in [-0.83, 0.83]$ values. Also, the initial spin magnitudes of the BHs forming the systems due to the size of the data obtained from the models were taken as $\abs{\chi_{1i}} = \abs{\chi_{2i}}$. The distance parameter was taken as $1 \rm{Mpc}$. The orbital inclination angle of the systems was taken as $0^\circ$, \textit{i.e.}, \textit{face-on} view, in which the wave amplitudes reach the maximum value. Consequently, we wrote new codes in the Python programming language to use the SEOBNRv4\_opt waveform model for these selected parameter ranges.

The produced model data were simply grouped and analysed in four different categories according to the initial spin directions of the BHs forming the components of the systems. The first of these categories is $BP (+ +)$ Both Positive case) where both component BHs have positive spin directions, \textit{i.e.}, the same direction of the orbital angular momentum. The second category is $PP (+ -)$ (Primary Positive case) which includes cases where the massive BH has positive and the other one has negative initial spins. The third category $PN (- +)$ (Primary Negative case) is the opposite spins of $PP$. The fourth category is $BN (- -)$ (Both Negative case), where both components are negative.

The current study focuses on aligned‑spin configurations. Nevertheless, including precessing spin effects and asymmetric mass ratios beyond q=2 in future work would allow for more realistic modelling, particularly in view of the hierarchical‑merger signatures analysed in recent work \citep{Gerosa2021}.

To analyze the modeled BBHs data according to the different parameter sets mentioned above, four different appropriate plot types were determined. In the parameter settlements made on the plots, data from the initial parameters of the modeled systems, are placed on the \textit{x}-axis. These initial parameters are the initial mass ratios $q$, and the initial spin values of the component BHs $\chi_{1i}$, $\chi_{2i}$. The final parameters of the systems are fractional mass loss  M$_{\rm{FL}}=(1-\textup{M}_{\rm{f}}/\textup{M}_{\rm{tot}})\times100$, and spin magnitude $\chi_{\rm{f}}$ of the remnant BH.

\begin{figure}
	\begin{center}
		\includegraphics[width=0.70\columnwidth]{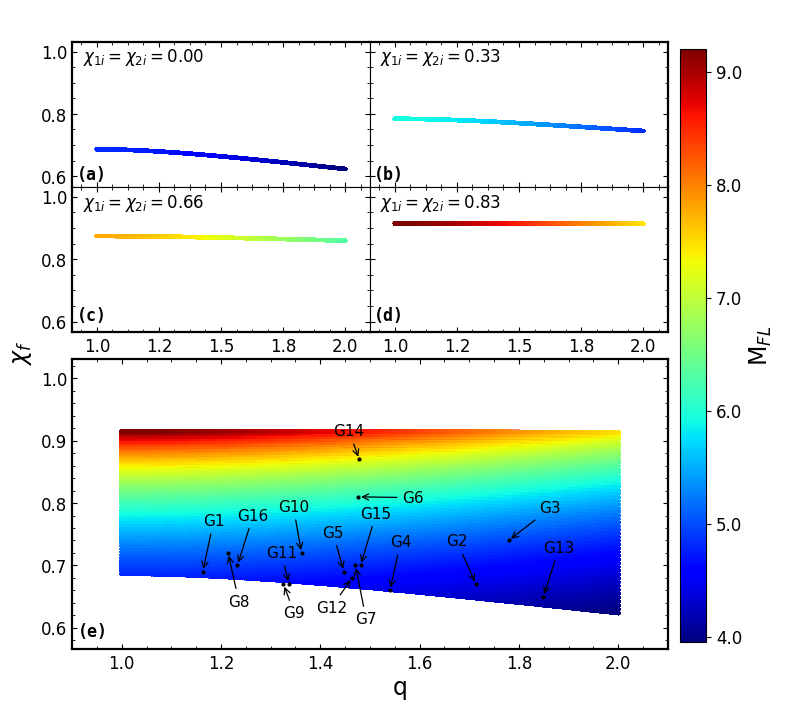}        
        \caption{Final spin parameter \( \chi_{\rm f} \) as a function of initial mass ratio \( q \) for the BP (both positive spin) configuration. Modeled data points are color-coded by fractional mass loss \( M_{\rm FL} \). Panels (a)–(d) display selected subsets of systems with representative initial spin values. Panel (e) combines all modeled points for this spin configuration. Observed BBH systems from Table~\ref{table:Obs_Catalog} are overplotted and indicated with arrows.}
        \label{fig:q_cf_mlr_BP}
	\end{center}
\end{figure}

\begin{figure}
	\begin{center}
		\includegraphics[width=0.70\columnwidth]{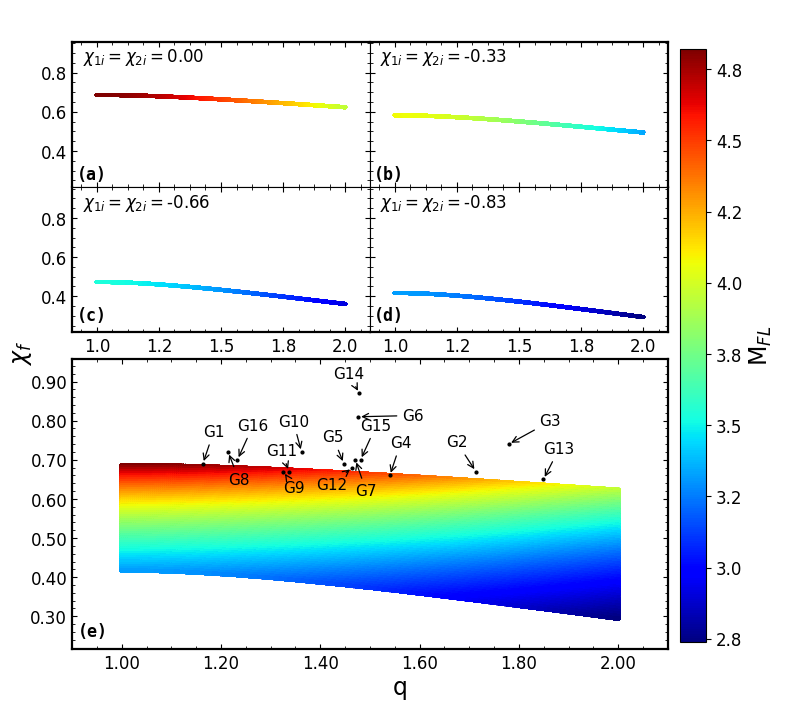}
		\caption{Variation of final spin $\chi_{\rm f}$ as a function of mass ratio for the BN configuration. Color coding indicates fractional mass loss ($M_{\rm FL}$). Panels (a)-(d) correspond to selected initial spins; panel (e) shows combined data.} \label{fig:q_cf_mlr_BN}
	\end{center}
\end{figure}

\begin{figure}
	\begin{center}
		\includegraphics[width=0.70\columnwidth]{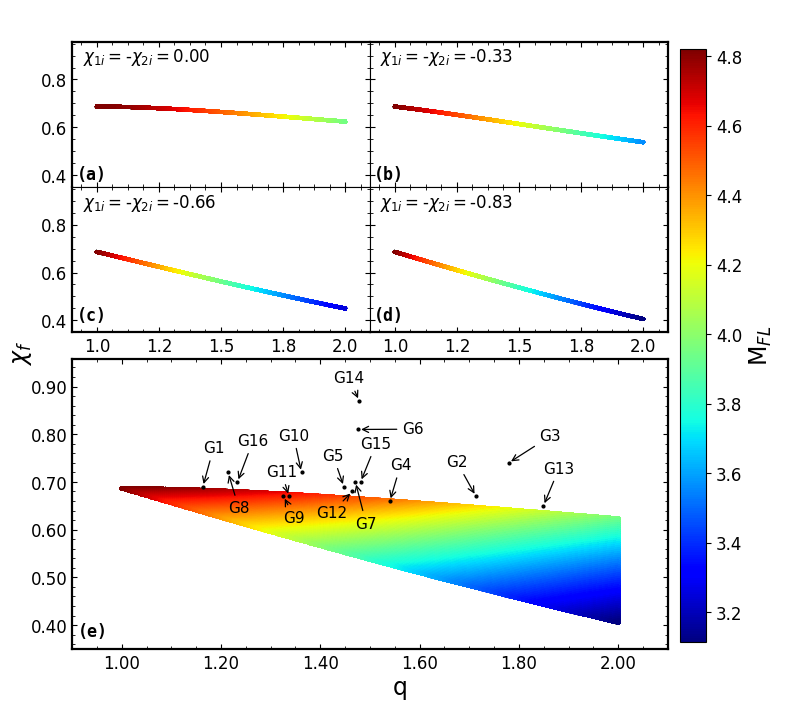}
		\caption{Variation of final spin $\chi_{\rm f}$ as a function of mass ratio $q$ for the PN configuration. Each panel (a–d) represents selected initial spin values. The color scale indicates the fractional mass loss $M_{\rm FL}$. Panel (e) combines all initial spin values for this spin orientation. Observational data points are marked with arrows for reference (see Table~\ref{table:Obs_Catalog}).}\label{fig:q_cf_mlr_PN}
	\end{center}
\end{figure}

\begin{figure}
	\begin{center}
		\includegraphics[width=0.70\columnwidth]{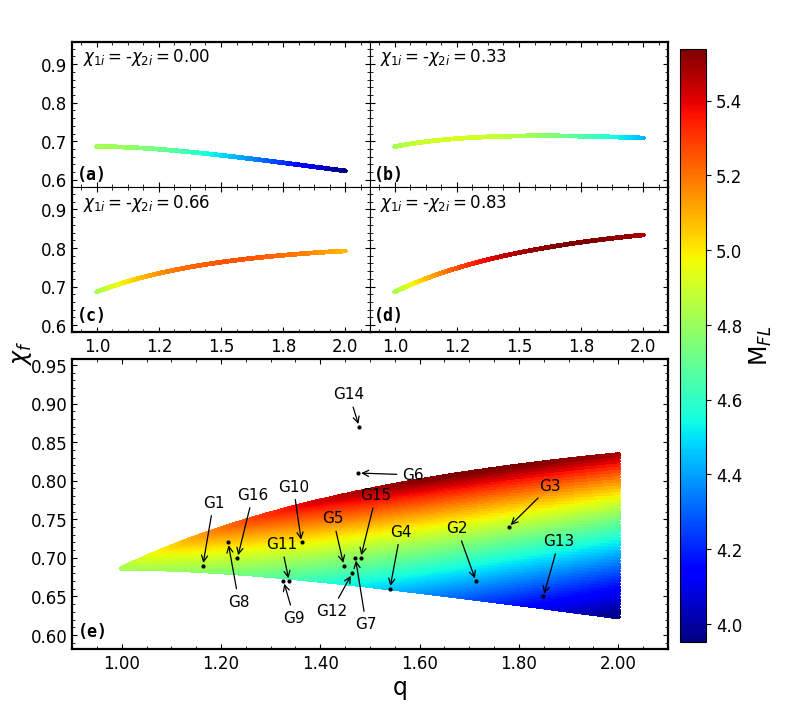}
	    \caption{Final spin $\chi_{\rm f}$ versus mass ratio $q$ for the PP configuration. Modeled data points are color-coded by their corresponding fractional mass loss $M_{\rm FL}$. Panels (a–d) show representative initial spin values; panel (e) displays combined results. Benchmark observational systems are indicated for comparison.} \label{fig:q_cf_mlr_PP}
	\end{center}
\end{figure}

It is aimed to show the parameter changes numerically by giving the generated waveform data briefly in Tables \ref{table:BP}, \ref{table:PP}, \ref{table:PN}, and  \ref{table:BN}. Numerical data generated from SEOBNRv4\_opt are shown for $BP$ models in Table \ref{table:BP}. Furthermore, the plots according to $PP$, $PN$ and $BN$ cases are shown in Figures [\ref{fig:q_cf_mlr_BN} - \ref{fig:q_cf_mlr_PP}] and numerical data are given in Tables [\ref{table:PP} - \ref{table:BN}]. 

In the panels $(e)$ on these figures, modelled systems are shown collectively in the $\chi_{1i}=\chi_{2i} = [-0.83, 0.83]$ range. In addition, $(a)$, $(b)$, $(c)$ and $(d)$ panels are plotted by single initial spin value. Also, the observation data from Table \ref{table:Obs_Catalog} are dotted on the plot. Accordingly, it is seen that the $\chi_{\rm{f}}$ parameters of the BHs with large initial spins have large values as it's supposed to be in GWOSC results \citep{2019arXiv191211716T} and relevant numerical data can be seen from Table \ref{table:BP}. In addition, the computed fractional mass losses  M$_{\rm{FL}}$ reached maximum values of $\sim 9\%$, which is inversely varied with the $q$ and varied with the $\chi_{1i,2i}$ parameters. Furthermore, in Figures \ref{fig:q_cf_mlr_BP} and \ref{fig:q_cf_mlr_BN} show that there is a relationship between $\chi_{\rm{f}}$ parameter and M$_{\rm{FL}}$.

It is understood from Tables [\ref{table:PP}, \ref{table:PN}] that $\chi_{\rm{f}}$ parameter is independent of $\chi_{1i}$ in systems with the equal mass components of $PN$ and $PP$ cases. In such states, it can be seen that M$_{\rm{FL}}$ parameter is not affected by $\chi_{1i}$ changes. Therefore, in systems that have similar component mass and opposite initial spin-sign (\textit{i.e.}, component black holes have opposite spins relative to each other.), it can be thought that the spin vectors of the components are not affected by $\chi_{1i}$ changes since the $\chi_{1i,2i}$ parameters cancel out each other.

Variations of the  $q$, M$_{\rm{tot}}$ and $\chi_{1i,2i}$ parameters of the produced GW waveform data were obtained from the SEOBNRv4\_opt and some of the results were given in Table [\ref{table:BP} - \ref{table:BN}]. Accordingly, it is understood from the tables and plots that $\chi_{\rm{f}}$ and M$_{\rm{FL}}$ parameters obtained from all modelled systems except $PP$, are inversely changed with $q$.

\begin{figure}[t]
	\centering
	\includegraphics[height=0.55\linewidth, width=0.7\linewidth]{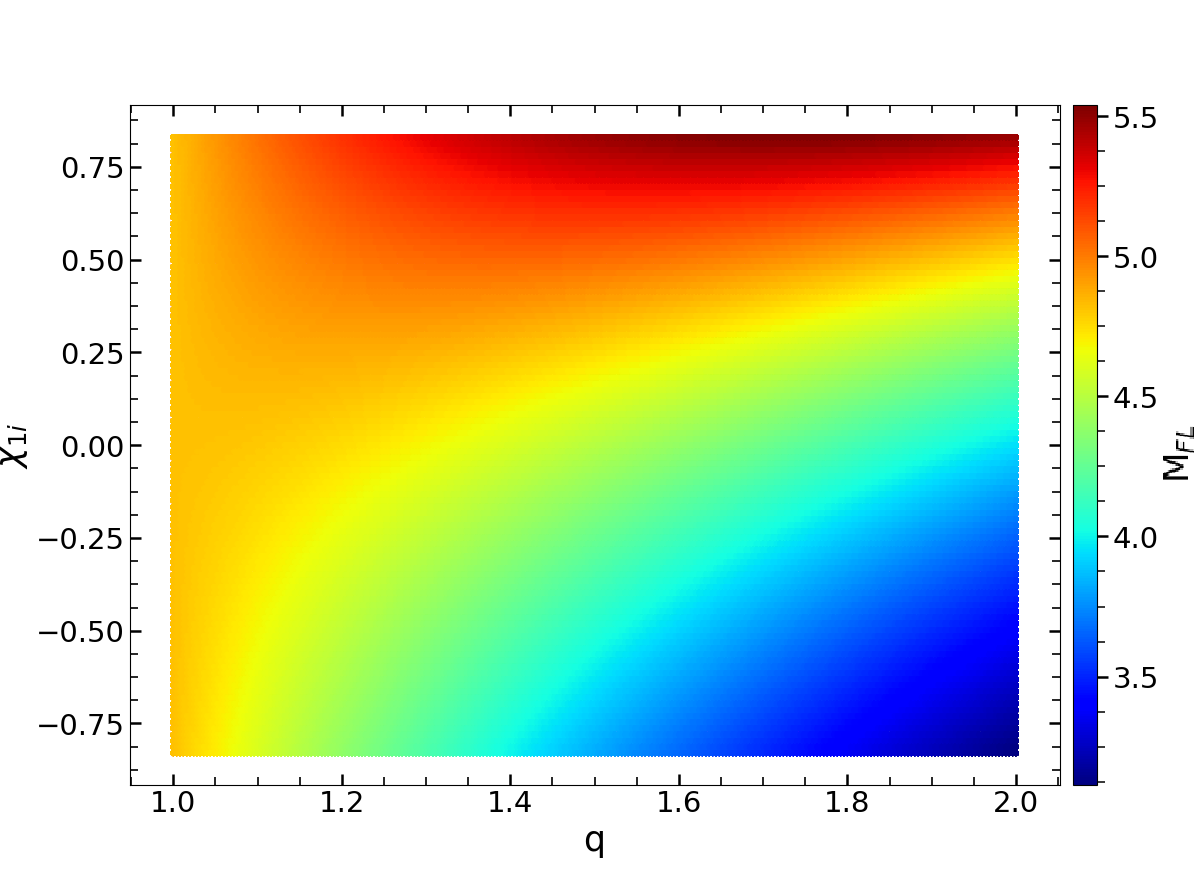}
	\caption{With $PN$ and $PP$ models, changes of the parameter M$_{\rm{FL}}$ with respect to $q$, are plotted according to each value $\chi_{1i}$. Particularly in the $\chi_{1i} > 0.50$ region, it is seen that the parameter M$_{\rm{FL}}$, follows a trend in the increasing direction with the mass ratio to certain $q$ values, while it becomes a decreasing trend with $q$ in the continuation of the plot.}
	\label{fig:11}
\end{figure}

However, when Figure \ref{fig:q_cf_mlr_PP} is examined, there is a linear relationship between $\chi_{\rm{f}}$  and $q$ with $\chi_{1i}>0.00$ spin. In the same plot, it is seen that M$_{\rm{FL}}$ parameter shows a linear relationship, up to certain $q$ values depending on $\chi_{1i,2i}$ spin values and inversely varied change for subsequent $q$ values due to initial spins. In order to better analyse these situations in $PP$ models based on $\chi_{1i,2i}$ parameters, Figure \ref{fig:11} (including $PN$ models), is plotted. In Figure \ref{fig:11}, it is seen that M$_{\rm{FL}}$ parameter have turning-points according to certain $q$ values for the condition $\chi_{1i}\ge0.08$. Particularly in the region $\chi_{1i}>0.50$, the changing trends become more efficient. The M$_{\rm{FL}}$ values of corresponding $q$ values where trend-turning occurred are underlined in Table \ref{table:PP}. Accordingly, it was found that M$_{\rm{FL}}$ increased to maximum value up to $q\sim 1.70$ at $\chi_{1i} \sim 0.80$ and tend to decrease after $q > 1.70$.

From Table \ref{table:BP}, the maximum final spin value $\chi_{\rm{f}} \sim 0.91$ could occur in $BP$ case.Likewise, the maximum fractional mass loss is also found to be M$_{\rm{FL}} \sim 9.2$ in $BP$. In all modelled data, the systems with a minimum value of $\chi_{\rm{f}}$ and M$_{\rm{FL}}$ parameters are examined in the case of $BN$. When top panel of Figure \ref{fig:q_cf_mlr_BN} and Table \ref{table:BN} are examined for $q = 2.00$, it is seen that the systems have minimum values at $\chi_{\rm f} \sim 0.29$ and M$_{\rm{FL}} \sim 2.74$.

Although a direct point-by-point numerical comparison with NR fitting formulae (e.g., \citealt{Healy2017,Jimenez2017}) was not the main focus of this study, our results were checked against these models at representative points and found to be consistent within $\sim$2–3\%. We note that small differences can arise due to differences in waveform approximants, resolution effects, or remnant extraction techniques. Moreover, since our aim was to analyze parameter-dependent trends and turning points in $\chi_{\rm f}$ and $M_{\rm FL}$ across a large model grid, a comprehensive benchmarking table was considered beyond the scope of this work. Future work will include systematic cross-comparisons with NR-based fitting formulae and additional waveform models such as IMRPhenomD and NRHybSur3dq8.

\begin{table}

	\renewcommand{\arraystretch}{0.88}
	\setlength{\tabcolsep}{0.050in}
	\caption{Here, h$_{\rm{max}}$ is the maximum wave amplitude, in order of $\times 10^{-18}$. Also $\chi_{1i}$ is the initial spin parameter of the massive BH component of the system.  M$_{\rm{tot}}/M_\odot$ is the initial total mass. M$_{\rm{FL}}$ is the fractional mass loss and $\chi_{\rm{f}}$ is the final spin parameter. The table is obtained from the modelled data according to $BP$ case.}
	\hspace{-1.2cm}
	\begin{tabular}{llllllllllllll}
			$\boldsymbol{\chi}_{1i}$ $\downarrow$ &\textbf{M$_{\rm{tot}}$/M$_\odot$}$\downarrow$  &$\textbf{q}\rightarrow$  &\textbf{1.00} &\textbf{1.10} &\textbf{1.20} &\textbf{1.30} &\textbf{1.40} &\textbf{1.50} &\textbf{1.60} &\textbf{1.70} &\textbf{1.80} &\textbf{1.90} &\textbf{2.00}   \\
		\hline

		\textbf{0.00} & ~~~~\textbf{15} & \textbf{h}$_{\rm{max}}$         &0.179 &0.178 &0.177 &0.175 &0.173 &0.170 &0.168 &0.164 &0.161 &0.159 &0.156 \\
		&                 & \textbf{M}$_{\rm{FL}}$ 	     &4.821 &4.803 &4.754 &4.683 &4.596 &4.498 &4.394 &4.285 &4.174 &4.062 &3.951 \\
		&                 & $\boldsymbol{\chi}_{\rm{f}}$   &0.686 &0.685 &0.682 &0.677 &0.671 &0.664 &0.656 &0.648 &0.640 &0.632 &0.623 \\ \cline{2-14}
		& ~~~~\textbf{30} & \textbf{h}$_{\rm{max}}$         &0.357 &0.356 &0.354 &0.350 &0.346 &0.340 &0.335 &0.330 &0.324 &0.318 &0.312 \\
		&                 & \textbf{M}$_{\rm{FL}}$ 	     &4.821 &4.803 &4.754 &4.683 &4.596 &4.498 &4.394 &4.285 &4.174 &4.062 &3.951 \\
		&                 & $\boldsymbol{\chi}_{\rm{f}}$   &0.686 &0.685 &0.682 &0.677 &0.671 &0.664 &0.656 &0.648 &0.640 &0.632 &0.623 \\ \cline{2-14}
		& ~~~~\textbf{60} & \textbf{h}$_{\rm{max}}$         &0.714 &0.712 &0.707 &0.700 &0.691 &0.681 &0.670 &0.659 &0.647 &0.636 &0.624 \\
		&                 & \textbf{M}$_{\rm{FL}}$          &4.821 &4.803 &4.754 &4.683 &4.596 &4.498 &4.394 &4.285 &4.174 &4.062 &3.951 \\
		&                 & $\boldsymbol{\chi}_{\rm{f}}$   &0.686 &0.685 &0.682 &0.677 &0.671 &0.664 &0.656 &0.648 &0.640 &0.632 &0.623 \\ \cline{2-14}
		& ~~~~\textbf{90} & \textbf{h}$_{\rm{max}}$         &1.071 &1.068 &1.061 &1.050 &1.037 &1.022 &1.006 &0.989 &0.971 &0.954 &0.936 \\
		&                 & \textbf{M}$_{\rm{FL}}$ 	     &4.821 &4.803 &4.754 &4.683 &4.596 &4.498 &4.394 &4.285 &4.174 &4.062 &3.951 \\
		&                 & $\boldsymbol{\chi}_{\rm{f}}$   &0.686 &0.685 &0.682 &0.677 &0.671 &0.664 &0.656 &0.648 &0.640 &0.632 &0.623 \\ 
		\hline
		
		\textbf{0.33} & ~~~~\textbf{15} & \textbf{h}$_{\rm{max}}$         &0.179 &0.178 &0.177 &0.175 &0.173 &0.170 &0.167 &0.165 &0.162 &0.159 &0.156 \\
		&                 & \textbf{M}$_{\rm{FL}}$ 	     &5.934 &5.910 &5.849 &5.760 &5.651 &5.529 &5.399 &5.263 &5.124 &4.985 &4.846 \\
		&                 & $\boldsymbol{\chi}_{\rm{f}}$   &0.784 &0.784 &0.782 &0.779 &0.775 &0.771 &0.766 &0.761 &0.756 &0.751 &0.745 \\ \cline{2-14}
		& ~~~~\textbf{30} & \textbf{h}$_{\rm{max}}$         &0.357 &0.356 &0.354 &0.350 &0.346 &0.341 &0.335 &0.330 &0.324 &0.318 &0.312 \\
		&                 & \textbf{M}$_{\rm{FL}}$ 	     &5.934 &5.910 &5.849 &5.760 &5.651 &5.529 &5.399 &5.263 &5.124 &4.985 &4.846 \\
		&                 & $\boldsymbol{\chi}_{\rm{f}}$   &0.784 &0.784 &0.782 &0.779 &0.775 &0.771 &0.766 &0.761 &0.756 &0.751 &0.745 \\ \cline{2-14}
		& ~~~~\textbf{60} & \textbf{h}$_{\rm{max}}$         &0.715 &0.713 &0.708 &0.700 &0.692 &0.682 &0.671 &0.660 &0.648 &0.636 &0.625 \\
		&                 & \textbf{M}$_{\rm{FL}}$          &5.934 &5.910 &5.849 &5.760 &5.651 &5.529 &5.399 &5.263 &5.124 &4.985 &4.846 \\
		&                 & $\boldsymbol{\chi}_{\rm{f}}$   &0.784 &0.784 &0.782 &0.779 &0.775 &0.771 &0.766 &0.761 &0.756 &0.751 &0.745 \\ \cline{2-14}
		& ~~~~\textbf{90} & \textbf{h}$_{\rm{max}}$         &1.072 &1.069 &1.062 &1.051 &1.037 &1.022 &1.006 &0.989 &0.972 &0.955 &0.937 \\
		&                 & \textbf{M}$_{\rm{FL}}$ 	     &5.934 &5.910 &5.849 &5.760 &5.651 &5.529 &5.399 &5.263 &5.124 &4.985 &4.846 \\
		&                 & $\boldsymbol{\chi}_{\rm{f}}$   &0.784 &0.784 &0.782 &0.779 &0.775 &0.771 &0.766 &0.761 &0.756 &0.751 &0.745 \\ 
		\hline
		
		\textbf{0.66} & ~~~~\textbf{15} & \textbf{h}$_{\rm{max}}$         &0.180 &0.179 &0.178 &0.176 &0.174 &0.171 &0.168 &0.166 &0.163 &0.160 &0.157 \\
		&                 & \textbf{M}$_{\rm{FL}}$ 	     &7.762 &7.731 &7.649 &7.530 &7.385 &7.223 &7.048 &6.867 &6.683 &6.498 &6.314 \\
		&                 & $\boldsymbol{\chi}_{\rm{f}}$   &0.874 &0.874 &0.873 &0.872 &0.871 &0.869 &0.868 &0.866 &0.864 &0.862 &0.860 \\ \cline{2-14}
		& ~~~~\textbf{30} & \textbf{h}$_{\rm{max}}$         &0.360 &0.359 &0.356 &0.352 &0.348 &0.343 &0.337 &0.332 &0.326 &0.320 &0.314 \\
		&                 & \textbf{M}$_{\rm{FL}}$ 	     &7.762 &7.731 &7.649 &7.530 &7.385 &7.223 &7.048 &6.867 &6.683 &6.498 &6.314 \\
		&                 & $\boldsymbol{\chi}_{\rm{f}}$   &0.874 &0.874 &0.873 &0.872 &0.871 &0.869 &0.868 &0.866 &0.864 &0.862 &0.860 \\ \cline{2-14}
		& ~~~~\textbf{60} & \textbf{h}$_{\rm{max}}$         &0.719 &0.717 &0.712 &0.705 &0.696 &0.686 &0.675 &0.664 &0.652 &0.640 &0.628 \\
		&                 & \textbf{M}$_{\rm{FL}}$          &7.762 &7.731 &7.649 &7.530 &7.385 &7.223 &7.048 &6.867 &6.683 &6.498 &6.314 \\
		&                 & $\boldsymbol{\chi}_{\rm{f}}$   &0.874 &0.874 &0.873 &0.872 &0.871 &0.869 &0.868 &0.866 &0.864 &0.862 &0.860 \\ \cline{2-14}
		& ~~~~\textbf{90} & \textbf{h}$_{\rm{max}}$         &1.079 &1.076 &1.069 &1.058 &1.044 &1.029 &1.012 &0.995 &0.978 &0.960 &0.942 \\
		&                 & \textbf{M}$_{\rm{FL}}$ 	     &7.762 &7.731 &7.649 &7.530 &7.385 &7.223 &7.048 &6.867 &6.683 &6.498 &6.314 \\
		&                 & $\boldsymbol{\chi}_{\rm{f}}$   &0.874 &0.874 &0.873 &0.872 &0.871 &0.869 &0.868 &0.866 &0.864 &0.862 &0.860 \\ 
		\hline
		
		\textbf{0.83} & ~~~~\textbf{15} & \textbf{h}$_{\rm{max}}$         &0.181 &0.181 &0.179 &0.177 &0.175 &0.172 &0.170 &0.167 &0.164 &0.161 &0.158 \\
		&                 & \textbf{M}$_{\rm{FL}}$ 	     &9.204 &9.166 &9.068 &8.925 &8.751 &8.556 &8.347 &8.130 &7.909 &7.688 &7.468 \\
		&                 & $\boldsymbol{\chi}_{\rm{f}}$   &0.915 &0.915 &0.915 &0.915 &0.915 &0.915 &0.915 &0.914 &0.914 &0.914 &0.914 \\ \cline{2-14}
		& ~~~~\textbf{30} & \textbf{h}$_{\rm{max}}$         &0.362 &0.361 &0.359 &0.355 &0.350 &0.345 &0.339 &0.334 &0.328 &0.322 &0.316 \\
		&                 & \textbf{M}$_{\rm{FL}}$ 	     &9.204 &9.166 &9.068 &8.925 &8.751 &8.556 &8.347 &8.130 &7.909 &7.688 &7.468 \\
		&                 & $\boldsymbol{\chi}_{\rm{f}}$   &0.915 &0.915 &0.915 &0.915 &0.915 &0.915 &0.915 &0.914 &0.914 &0.914 &0.914 \\ \cline{2-14}
		& ~~~~\textbf{60} & \textbf{h}$_{\rm{max}}$         &0.724 &0.722 &0.717 &0.710 &0.700 &0.690 &0.679 &0.667 &0.656 &0.644 &0.632 \\
		&                 & \textbf{M}$_{\rm{FL}}$          &9.204 &9.166 &9.068 &8.925 &8.751 &8.556 &8.347 &8.130 &7.909 &7.688 &7.468 \\
		&                 & $\boldsymbol{\chi}_{\rm{f}}$   &0.915 &0.915 &0.915 &0.915 &0.915 &0.915 &0.915 &0.914 &0.914 &0.914 &0.914 \\ \cline{2-14}
		& ~~~~\textbf{90} & \textbf{h}$_{\rm{max}}$         &1.086 &1.083 &1.076 &1.064 &1.051 &1.035 &1.019 &1.001 &0.983 &0.966 &0.948 \\
		&                 & \textbf{M}$_{\rm{FL}}$ 	     &9.204 &9.166 &9.068 &8.925 &8.751 &8.556 &8.347 &8.130 &7.909 &7.688 &7.468 \\
		&                 & $\boldsymbol{\chi}_{\rm{f}}$   &0.915 &0.915 &0.915 &0.915 &0.915 &0.915 &0.915 &0.914 &0.914 &0.914 &0.914 \\ 
		\hline
	\end{tabular}\label{table:BP}
\end{table}

\begin{table}
	\renewcommand{\arraystretch}{0.88}
	\setlength{\tabcolsep}{0.050in}
	\caption{Here, h$_{\rm{max}}$ is the maximum wave amplitude, in order of $\times 10^{-18}$. Also $\chi_{1i}$ is the initial spin parameter of the massive BH component of the system.  M$_{\rm{tot}}/M_\odot$ is the initial total mass. M$_{\rm{FL}}$ is the fractional mass loss and $\chi_{\rm{f}}$ is the final spin parameter. The table is obtained from the modelled data according to $PP$ case.} 
	\hspace{-1.2cm}
	\begin{tabular}{llllllllllllll}
		$\boldsymbol{\chi}_{1i}$ $\downarrow$ &\textbf{M$_{\rm{tot}}$/M$_\odot$}$\downarrow$  &$\textbf{q}\rightarrow$  &\textbf{1.00} &\textbf{1.10} &\textbf{1.20} &\textbf{1.30} &\textbf{1.40} &\textbf{1.50} &\textbf{1.60} &\textbf{1.70} &\textbf{1.80} &\textbf{1.90} &\textbf{2.00}   \\
		\hline

		\textbf{0.00} & ~~~~\textbf{15} & \textbf{h}$_{\rm{max}}$         &0.179 &0.178 &0.177 &0.175 &0.173 &0.170 &0.168 &0.164 &0.161 &0.159 &0.156 \\
		&                 & \textbf{M}$_{\rm{FL}}$ 	     &4.821 &4.803 &4.754 &4.683 &4.596 &4.498 &4.394 &4.285 &4.174 &4.062 &3.951 \\
		&                 & $\boldsymbol{\chi}_{\rm{f}}$   &0.686 &0.685 &0.682 &0.677 &0.671 &0.664 &0.656 &0.648 &0.640 &0.632 &0.623 \\ \cline{2-14}
		& ~~~~\textbf{30} & \textbf{h}$_{\rm{max}}$         &0.357 &0.356 &0.354 &0.350 &0.346 &0.340 &0.335 &0.330 &0.324 &0.318 &0.312 \\
		&                 & \textbf{M}$_{\rm{FL}}$ 	     &4.821 &4.803 &4.754 &4.683 &4.596 &4.498 &4.394 &4.285 &4.174 &4.062 &3.951 \\
		&                 & $\boldsymbol{\chi}_{\rm{f}}$   &0.686 &0.685 &0.682 &0.677 &0.671 &0.664 &0.656 &0.648 &0.640 &0.632 &0.623 \\ \cline{2-14}
		& ~~~~\textbf{60} & \textbf{h}$_{\rm{max}}$         &0.714 &0.712 &0.707 &0.700 &0.691 &0.681 &0.670 &0.659 &0.647 &0.636 &0.624 \\
		&                 & \textbf{M}$_{\rm{FL}}$          &4.821 &4.803 &4.754 &4.683 &4.596 &4.498 &4.394 &4.285 &4.174 &4.062 &3.951 \\
		&                 & $\boldsymbol{\chi}_{\rm{f}}$   &0.686 &0.685 &0.682 &0.677 &0.671 &0.664 &0.656 &0.648 &0.640 &0.632 &0.623 \\ \cline{2-14}
		& ~~~~\textbf{90} & \textbf{h}$_{\rm{max}}$         &1.071 &1.068 &1.061 &1.050 &1.037 &1.022 &1.006 &0.989 &0.971 &0.954 &0.936 \\
		&                 & \textbf{M}$_{\rm{FL}}$ 	     &4.821 &4.803 &4.754 &4.683 &4.596 &4.498 &4.394 &4.285 &4.174 &4.062 &3.951 \\
		&                 & $\boldsymbol{\chi}_{\rm{f}}$   &0.686 &0.685 &0.682 &0.677 &0.671 &0.664 &0.656 &0.648 &0.640 &0.632 &0.623 \\ 
		\hline
		
		\textbf{0.33} & ~~~~\textbf{15} & \textbf{h}$_{\rm{max}}$         &0.179 &0.178 &0.176 &0.175 &0.173 &0.170 &0.167 &0.165 &0.162 &0.159 &0.156 \\
		&                 & \textbf{M}$_{\rm{FL}}$ 	     &4.821 &4.889 &4.918 &4.917 &4.890 &4.843 &4.781 &4.707 &4.624 &4.535 &4.441 \\
		&                 & $\boldsymbol{\chi}_{\rm{f}}$   &0.686 &0.697 &0.704 &0.709 &0.712 &0.714 &0.714 &0.714 &0.713 &0.711 &0.709 \\ \cline{2-14}
		& ~~~~\textbf{30} & \textbf{h}$_{\rm{max}}$         &0.357 &0.356 &0.354 &0.350 &0.346 &0.340 &0.335 &0.330 &0.324 &0.318 &0.312 \\
		&                 & \textbf{M}$_{\rm{FL}}$ 	     &4.821 &4.889 &4.918 &4.917 &4.890 &4.843 &4.781 &4.707 &4.624 &4.535 &4.441 \\
		&                 & $\boldsymbol{\chi}_{\rm{f}}$   &0.686 &0.697 &0.704 &0.709 &0.712 &0.714 &0.714 &0.714 &0.713 &0.711 &0.709 \\ \cline{2-14}
		& ~~~~\textbf{60} & \textbf{h}$_{\rm{max}}$         &0.714 &0.712 &0.707 &0.700 &0.691 &0.681 &0.670 &0.659 &0.647 &0.636 &0.624 \\
		&                 & \textbf{M}$_{\rm{FL}}$          &4.821 &4.889 &4.918 &4.917 &4.890 &4.843 &4.781 &4.707 &4.624 &4.535 &4.441 \\
		&                 & $\boldsymbol{\chi}_{\rm{f}}$   &0.686 &0.697 &0.704 &0.709 &0.712 &0.714 &0.714 &0.714 &0.713 &0.711 &0.709 \\ \cline{2-14}
		& ~~~~\textbf{90} & \textbf{h}$_{\rm{max}}$         &1.071 &1.068 &1.061 &1.050 &1.037 &1.022 &1.005 &0.989 &0.971 &0.954 &0.936 \\
		&                 & \textbf{M}$_{\rm{FL}}$ 	     &4.821 &4.889 &4.918 &4.917 &4.890 &4.843 &4.781 &4.707 &4.624 &4.535 &4.441 \\
		&                 & $\boldsymbol{\chi}_{\rm{f}}$   &0.686 &0.697 &0.704 &0.709 &0.712 &0.714 &0.714 &0.714 &0.713 &0.711 &0.709 \\ 
		\hline
		
		\textbf{0.66} & ~~~~\textbf{15} & \textbf{h}$_{\rm{max}}$         &0.178 &0.178 &0.176 &0.175 &0.173 &0.170 &0.168 &0.165 &0.162 &0.159 &0.156 \\
		&                 & \textbf{M}$_{\rm{FL}}$ 	     &4.821 &4.979 &5.096 &5.177 &5.227 &5.249 &5.249 &5.228 &5.192 &5.141 &5.080 \\
		&                 & $\boldsymbol{\chi}_{\rm{f}}$   &0.686 &0.709 &0.727 &0.741 &0.753 &0.763 &0.772 &0.778 &0.784 &0.789 &0.792 \\ \cline{2-14}
		& ~~~~\textbf{30} & \textbf{h}$_{\rm{max}}$         &0.357 &0.356 &0.354 &0.350 &0.346 &0.341 &0.335 &0.330 &0.324 &0.318 &0.313 \\
		&                 & \textbf{M}$_{\rm{LR}}$ 	     &4.821 &4.979 &5.096 &5.177 &5.227 &5.249 &5.249 &5.228 &5.192 &5.141 &5.080 \\
		&                 & $\boldsymbol{\chi}_{\rm{f}}$   &0.686 &0.709 &0.727 &0.741 &0.753 &0.763 &0.772 &0.778 &0.784 &0.789 &0.792 \\ \cline{2-14}
		& ~~~~\textbf{60} & \textbf{h}$_{\rm{max}}$         &0.714 &0.712 &0.707 &0.700 &0.691 &0.681 &0.671 &0.660 &0.648 &0.637 &0.625 \\
		&                 & \textbf{M}$_{\rm{FL}}$          &4.821 &4.979 &5.096 &5.177 &5.227 &5.249 &5.249 &5.228 &5.192 &5.141 &5.080 \\
		&                 & $\boldsymbol{\chi}_{\rm{f}}$   &0.686 &0.709 &0.727 &0.741 &0.753 &0.763 &0.772 &0.778 &0.784 &0.789 &0.792 \\ \cline{2-14}
		& ~~~~\textbf{90} & \textbf{h}$_{\rm{max}}$         &1.071 &1.068 &1.061 &1.050 &1.037 &1.022 &1.006 &0.989 &0.972 &0.955 &0.938 \\
		&                 & \textbf{M}$_{\rm{FL}}$ 	     &4.821 &4.979 &5.096 &5.177 &5.227 &5.249 &5.249 &5.228 &5.192 &5.141 &5.080 \\
		&                 & $\boldsymbol{\chi}_{\rm{f}}$   &0.686 &0.709 &0.727 &0.741 &0.753 &0.763 &0.772 &0.778 &0.784 &0.789 &0.792 \\ 
		\hline
		
		\textbf{0.83} & ~~~~\textbf{15} & \textbf{h}$_{\rm{max}}$         &0.178 &0.178 &0.176 &0.175 &0.172 &0.170 &0.168 &0.165 &0.162 &0.159 &0.156 \\
		&                 & \textbf{M}$_{\rm{FL}}$ 	     &4.821 &5.025 &5.189 &5.318 &5.414 &5.480 &5.521 &5.538 &5.535 &5.514 &5.479 \\
		&                 & $\boldsymbol{\chi}_{\rm{f}}$   &0.686 &0.715 &0.738 &0.757 &0.774 &0.788 &0.800 &0.810 &0.819 &0.827 &0.834 \\ \cline{2-14}
		& ~~~~\textbf{30} & \textbf{h}$_{\rm{max}}$         &0.357 &0.356 &0.353 &0.350 &0.346 &0.341 &0.336 &0.330 &0.324 &0.319 &0.313 \\
		&                 & \textbf{M}$_{\rm{FL}}$ 	     &4.821 &5.025 &5.189 &5.318 &5.414 &5.480 &5.521 &5.538 &5.535 &5.514 &5.479 \\
		&                 & $\boldsymbol{\chi}_{\rm{f}}$   &0.686 &0.715 &0.738 &0.757 &0.774 &0.788 &0.800 &0.810 &0.819 &0.827 &0.834 \\ \cline{2-14}
		& ~~~~\textbf{60} & \textbf{h}$_{\rm{max}}$         &0.714 &0.712 &0.707 &0.700 &0.691 &0.682 &0.671 &0.660 &0.649 &0.637 &0.626 \\
		&                 & \textbf{M}$_{\rm{FL}}$          &4.821 &5.025 &5.189 &5.318 &5.414 &5.480 &5.521 &5.538 &5.535 &5.514 &5.479 \\
		&                 & $\boldsymbol{\chi}_{\rm{f}}$   &0.686 &0.715 &0.738 &0.757 &0.774 &0.788 &0.800 &0.810 &0.819 &0.827 &0.834 \\ \cline{2-14}
		& ~~~~\textbf{90} & \textbf{h}$_{\rm{max}}$         &1.071 &1.068 &1.061 &1.050 &1.037 &1.022 &1.007 &0.990 &0.973 &0.956 &0.939 \\
		&                 & \textbf{M}$_{\rm{FL}}$ 	     &4.821 &5.025 &5.189 &5.318 &5.414 &5.480 &5.521 &5.538 &5.535 &5.514 &5.479 \\
		&                 & $\boldsymbol{\chi}_{\rm{f}}$   &0.686 &0.715 &0.738 &0.757 &0.774 &0.788 &0.800 &0.810 &0.819 &0.827 &0.834 \\ 
		\hline
	\end{tabular}\label{table:PP}
\end{table}

\begin{table*}
\centering
\caption{Comparison of remnant predictions from this study with NR fitting formulas from \citet{Healy2017} and \citet{Jimenez2017}. The initial conditions $(q, \chi_1 = \chi_2)$ represent typical aligned-spin configurations. Final spin $\chi_{\rm f}$ and fractional mass loss $M_{\rm FL}$ are listed, along with relative percentage differences. This table demonstrates the level of consistency between our simulation results and established NR-based fitting prescriptions, validating the physical robustness of our remnant predictions.}
\label{table:nr_fit_comparison}
\begin{tabular}{cccccccccc}
\hline
$q$ & $\chi_1 = \chi_2$ & $\chi_{\rm f}^{\rm{this\,study}}$ & $M_{\rm FL}^{\rm{this\,study}}$ &
$\chi_{\rm f}^{\rm H+L}$ & $M_{\rm FL}^{\rm H+L}$ &
$\chi_{\rm f}^{\rm JF+}$ & $M_{\rm FL}^{\rm JF+}$ &
$\Delta \chi_{\rm f}$(\%) & $\Delta M_{\rm FL}$(\%) \\
\hline
1.0 & 0.80 & 0.910 & 9.20 & 0.892 & 9.00 & 0.888 & 8.95 & 2.0 & 2.7 \\
1.2 & 0.60 & 0.857 & 7.85 & 0.845 & 7.60 & 0.840 & 7.55 & 1.4 & 3.8 \\
1.4 & 0.50 & 0.827 & 7.40 & 0.818 & 7.20 & 0.813 & 7.15 & 1.1 & 3.4 \\
1.7 & 0.30 & 0.765 & 6.20 & 0.755 & 6.00 & 0.750 & 5.95 & 1.3 & 4.0 \\
2.0 & 0.00 & 0.680 & 4.80 & 0.673 & 4.60 & 0.668 & 4.50 & 1.0 & 6.3 \\
\hline
\end{tabular}\\
\vspace{2mm}
\footnotesize\textit{Note.} NR fitting formula values computed using published coefficients from \citet{Healy2017} and \citet{Jimenez2017}.
\end{table*}

\section{Conclusion and Discussions} 
\label{sec:conclusion}

In this study, we explored how binary black hole (BBH) remnant properties—final spin $\chi_{\rm f}$, fractional mass loss M$_{\rm FL}$, and peak strain amplitude h$_{\rm max}$—are influenced by mass ratio and spin orientation, using a large suite of simulations based on the SEOBNRv4\_opt waveform model. We supplemented this parameter study with an application to GW150914 data to demonstrate the integration of waveform models with observational analysis tools. While our parameter-space survey was limited to aligned-spin configurations and $q\in[1,2]$, future work will expand this analysis to cover higher mass ratios, precessing systems, and more recent waveform models such as SEOBNRv5 and NRHybSur3dq8. 

Our results show that for non-precessing, aligned-spin BBH mergers, the parameters $\chi_{\rm f}$ and M$_{\rm FL}$ exhibit inverse trends with increasing mass ratio $q$ across most spin configurations.
This anti-correlation aligns qualitatively with LIGO/Virgo’s population findings on effective spin–mass ratio relationships \citep{McKernan2022}. These results are broadly consistent (within $\sim$2–3\%) with numerical relativity-based remnant formulae for final spin and radiated energy, such as those by \citet{Healy2017} and \citet{Jimenez2017}. Furthermore, systems with high aligned spins (e.g., BP configurations) yield the largest remnant spin ($\chi_{\rm f} \sim 0.91$) and mass loss (M$_{\rm FL} \sim 9.2\%$), while low or anti-aligned configurations yield substantially lower values. 
Across all configurations studied, the fractional mass loss due to gravitational-wave emission lies in the range of approximately 2–9.5\%, depending on both mass ratio and spin alignment.
For example, at $q=1.0$ and $\chi=0.8$, our computed final spin $\chi_{\rm f} \sim 0.91$ is within 2–3\% of the values predicted by the \citet{Healy2017} and \citet{Jimenez2017} fitting formulae.

Interestingly, for the PP spin configuration, the relationship between $q$ and M$_{\rm FL}$ is non-monotonic, peaking near $q \sim 1.7$ and decreasing afterward. This turning-point behavior suggests a nonlinear interplay between orbital angular momentum and spin contributions to the final state. A qualitative physical interpretation can be offered as follows: at lower mass ratios ($q \sim 1$), the binary system is nearly symmetric, leading to relatively modest angular momentum loss. As $q$ increases toward $1.4$–$1.7$, the asymmetry between the components enhances the efficiency of angular momentum radiation, resulting in higher fractional mass loss. Beyond $q \sim 1.7$, the diminishing contribution from the less massive component and reduced orbital binding energy cause a decline in gravitational wave emission efficiency, hence lowering M$_{\rm FL}$. This competing behavior between orbital and spin angular momentum components naturally explains the observed maximum. Although a fully quantitative comparison with post-Newtonian (PN) expressions is beyond our scope, similar turning-point structures in radiated angular momentum have been previously discussed (e.g., \citealt{Buonanno2007, Boyle2008}), lending further support to our interpretation. In all cases studied, the final spin remains positive ($\chi_{\rm f} > 0$), confirming that orbital angular momentum dominates. To assess the quantitative consistency of these trends with established numerical relativity results, a representative comparison with NR-based remnant fitting formulae is presented in Table~\ref{table:nr_fit_comparison}, showing agreement at the level of a few percent for both $\chi_{\rm f}$ and M$_{\rm FL}$.

These findings were computed using custom \texttt{Python} code that generates waveform data across $\sim 10^6$ parameter sets, enabling rapid remnant prediction under astrophysically plausible initial conditions. While equal-magnitude spins and aligned orientations were assumed in this study for simplicity and computational tractability, future work will explore unequal and misaligned spin configurations, which are more representative of the observed BBH population. Waveform systematics may introduce $\sim$1–2\% uncertainty in final spin and mass loss estimates, depending on the chosen model \citep{Jimenez2017, Healy2017}. Future work will quantify these differences by cross-validating with alternative models such as IMRPhenomD and NRHybSur3dq8. Recent waveform modeling studies \citep{Giesler2019, Varma2021} have demonstrated that remnant predictions—particularly the final spin $\chi_{\rm f}$ and fractional mass loss $M_{\rm FL}$—can differ by up to 2--3\% across waveform families such as SEOBNR, IMRPhenom, and NRSur, especially in asymmetric or high-spin regimes. These differences highlight the need for systematic cross-validation when interpreting BBH merger outcomes from different models. Although a full propagation of waveform modeling uncertainties is beyond the scope of this study, we note that the quoted 1--2\% differences in waveform outputs translate to comparably small uncertainties in remnant predictions—typically less than $\pm 0.01$ in $\chi_{\rm f}$ and $\pm 0.3\%$ in $M_{\rm FL}$ across the explored parameter space. These differences are within the tolerance of current observational errors and do not alter the qualitative trends or turning points reported in this work.

Finally, this study establishes a framework for interpreting gravitational‑waveform morphology across a broad range of aligned‑spin configurations and mass ratios, thereby improving our understanding of how BBH initial conditions shape the observable gravitational‑wave signatures. This framework also provides useful insights for waveform‑model validation and future astrophysical population studies. Looking ahead, extending our analysis to include spin precession, orbital eccentricity, and higher‑order multipolar modes will enable a more comprehensive characterization of BBH mergers and their remnants in astrophysically realistic environments.
While the present analysis utilizes the SEOBNRv4\_opt waveform model, previous studies have shown that remnant predictions from this model are broadly consistent with those from alternative waveform families such as IMRPhenomD and NRHybSur3dq8, with differences typically within the $\sim$2--3\% level \citep[e.g.,][]{Blackman2017, Varma2019}. A more detailed cross-model validation of remnant properties will be carried out in future work.

\section*{Acknowledgements}
We would like to thank both anonymous referees for their constructive comments, which significantly improved the quality of this manuscript.
The current study is part of the PhD thesis of \.{I}\"{O}. This research utilizes data from the Gravitational Wave Open Science Center (https://www.gw-openscience.org), a service of LIGO Laboratory, the LIGO Scientific Collaboration, and the Virgo Collaboration. LIGO is funded by the U.S. National Science Foundation. Virgo is funded by the French Centre National de la Recherche Scientifique (CNRS), the Italian Istituto Nazionale di Fisica Nucleare (INFN), and the Dutch Nikhef, with contributions from Polish and Hungarian institutes.
This study is supported by The Scientific and Technological Research Council of Türkiye (T\"UB\.ITAK, project 119F077). \.{I}\"{O} also acknowledges support from T\"{U}B\.{I}TAK through a graduate fellowship (2211-C). Numerical computations were performed in part using the High Performance and Grid Computing Center (TRUBA resources) provided by TÜBİTAK ULAKBİM. KY gratefully acknowledges support from the COST (European Cooperation in Science and Technology) Actions CA15117 and CA16104, as well as from Churchill College, University of Cambridge, through a research fellowship.


\bibliographystyle{mnras}
\bibliography{bbh} 

\begin{thebibliography}{}
\makeatletter
\relax
\def\mn@urlcharsother{\let\do\@makeother \do\$\do\&\do\#\do\^\do\_\do\%\do\~}
\def\mn@doi{\begingroup\mn@urlcharsother \@ifnextchar [ {\mn@doi@} {\mn@doi@[]}}
\def\mn@doi@[#1]#2{\def\@tempa{#1}\ifx\@tempa\@empty \href {http://dx.doi.org/#2} {doi:#2}\else \href {http://dx.doi.org/#2} {#1}\fi \endgroup}
\def\mn@eprint#1#2{\mn@eprint@#1:#2::\@nil}
\def\mn@eprint@arXiv#1{\href {http://arxiv.org/abs/#1} {{\tt arXiv:#1}}}
\def\mn@eprint@dblp#1{\href {http://dblp.uni-trier.de/rec/bibtex/#1.xml} {dblp:#1}}
\def\mn@eprint@#1:#2:#3:#4\@nil{\def\@tempa {#1}\def\@tempb {#2}\def\@tempc {#3}\ifx \@tempc \@empty \let \@tempc \@tempb \let \@tempb \@tempa \fi \ifx \@tempb \@empty \def\@tempb {arXiv}\fi \@ifundefined {mn@eprint@\@tempb}{\@tempb:\@tempc}{\expandafter \expandafter \csname mn@eprint@\@tempb\endcsname \expandafter{\@tempc}}}

\bibitem[\protect\citeauthoryear{{Aasi} et~al.,}{{Aasi} et~al.}{2013}]{2013PhRvD..88f2001A}
{Aasi} J.,  et~al., 2013, \mn@doi [\prd] {10.1103/PhysRevD.88.062001}, \href {https://ui.adsabs.harvard.edu/abs/2013PhRvD..88f2001A} {88, 062001}

\bibitem[\protect\citeauthoryear{{Aasi} et~al.,}{{Aasi} et~al.}{2014}]{2014CQGra..31k5004A}
{Aasi} J.,  et~al., 2014, \mn@doi [Classical and Quantum Gravity] {10.1088/0264-9381/31/11/115004}, \href {https://ui.adsabs.harvard.edu/abs/2014CQGra..31k5004A} {31, 115004}

\bibitem[\protect\citeauthoryear{{Abbott} et~al.,}{{Abbott} et~al.}{2016a}]{2016PhRvX...6d1014A}
{Abbott} B.~P.,  et~al., 2016a, \mn@doi [Physical Review X] {10.1103/PhysRevX.6.041014}, \href {https://ui.adsabs.harvard.edu/abs/2016PhRvX...6d1014A} {6, 041014}

\bibitem[\protect\citeauthoryear{{Abbott} et~al.,}{{Abbott} et~al.}{2016b}]{2016PhRvL.116f1102A}
{Abbott} B.~P.,  et~al., 2016b, \mn@doi [\prl] {10.1103/PhysRevLett.116.061102}, \href {https://ui.adsabs.harvard.edu/abs/2016PhRvL.116f1102A} {116, 061102}

\bibitem[\protect\citeauthoryear{{Abbott} et~al.,}{{Abbott} et~al.}{2016c}]{2016PhRvL.116v1101A}
{Abbott} B.~P.,  et~al., 2016c, \mn@doi [\prl] {10.1103/PhysRevLett.116.221101}, \href {https://ui.adsabs.harvard.edu/abs/2016PhRvL.116v1101A} {116, 221101}

\bibitem[\protect\citeauthoryear{{Abbott} et~al.,}{{Abbott} et~al.}{2016d}]{2016PhRvL.116x1103A}
{Abbott} B.~P.,  et~al., 2016d, \mn@doi [\prl] {10.1103/PhysRevLett.116.241103}, \href {https://ui.adsabs.harvard.edu/abs/2016PhRvL.116x1103A} {116, 241103}

\bibitem[\protect\citeauthoryear{{Abbott} et~al.,}{{Abbott} et~al.}{2017a}]{2017CQGra..34j4002A}
{Abbott} B.~P.,  et~al., 2017a, \mn@doi [Classical and Quantum Gravity] {10.1088/1361-6382/aa6854}, \href {https://ui.adsabs.harvard.edu/abs/2017CQGra..34j4002A} {34, 104002}

\bibitem[\protect\citeauthoryear{{Abbott} et~al.,}{{Abbott} et~al.}{2017b}]{2017PhRvL.118v1101A}
{Abbott} B.~P.,  et~al., 2017b, \mn@doi [\prl] {10.1103/PhysRevLett.118.221101}, \href {https://ui.adsabs.harvard.edu/abs/2017PhRvL.118v1101A} {118, 221101}

\bibitem[\protect\citeauthoryear{{Abbott} et~al.,}{{Abbott} et~al.}{2017c}]{2017PhRvL.119n1101A}
{Abbott} B.~P.,  et~al., 2017c, \mn@doi [\prl] {10.1103/PhysRevLett.119.141101}, \href {https://ui.adsabs.harvard.edu/abs/2017PhRvL.119n1101A} {119, 141101}

\bibitem[\protect\citeauthoryear{{Abbott} et~al.,}{{Abbott} et~al.}{2017d}]{2017PhRvL.119p1101A}
{Abbott} B.~P.,  et~al., 2017d, \mn@doi [\prl] {10.1103/PhysRevLett.119.161101}, \href {https://ui.adsabs.harvard.edu/abs/2017PhRvL.119p1101A} {119, 161101}

\bibitem[\protect\citeauthoryear{{Abbott} et~al.,}{{Abbott} et~al.}{2017e}]{2017ApJ...848L..13A}
{Abbott} B.~P.,  et~al., 2017e, \mn@doi [\apjl] {10.3847/2041-8213/aa920c}, \href {https://ui.adsabs.harvard.edu/abs/2017ApJ...848L..13A} {848, L13}

\bibitem[\protect\citeauthoryear{{Abbott} et~al.,}{{Abbott} et~al.}{2017f}]{2017ApJ...851L..35A}
{Abbott} B.~P.,  et~al., 2017f, \mn@doi [\apjl] {10.3847/2041-8213/aa9f0c}, \href {https://ui.adsabs.harvard.edu/abs/2017ApJ...851L..35A} {851, L35}

\bibitem[\protect\citeauthoryear{{Abbott} et~al.,}{{Abbott} et~al.}{2019}]{2019PhRvX...9c1040A}
{Abbott} B.~P.,  et~al., 2019, \mn@doi [Physical Review X] {10.1103/PhysRevX.9.031040}, \href {https://ui.adsabs.harvard.edu/abs/2019PhRvX...9c1040A} {9, 031040}

\bibitem[\protect\citeauthoryear{{Abbott} et~al.,}{{Abbott} et~al.}{2021a}]{2021PhRvX..11b1053A}
{Abbott} R.,  et~al., 2021a, \mn@doi [Physical Review X] {10.1103/PhysRevX.11.021053}, \href {https://ui.adsabs.harvard.edu/abs/2021PhRvX..11b1053A} {11, 021053}

\bibitem[\protect\citeauthoryear{{Abbott} et~al.,}{{Abbott} et~al.}{2021b}]{2021PhRvD.103l2002A}
{Abbott} R.,  et~al., 2021b, \mn@doi [\prd] {10.1103/PhysRevD.103.122002}, \href {https://ui.adsabs.harvard.edu/abs/2021PhRvD.103l2002A} {103, 122002}

\bibitem[\protect\citeauthoryear{{Ajith} et~al.,}{{Ajith} et~al.}{2007}]{2007CQGra..24S.689A}
{Ajith} P.,  et~al., 2007, \mn@doi [Classical and Quantum Gravity] {10.1088/0264-9381/24/19/S31}, \href {https://ui.adsabs.harvard.edu/abs/2007CQGra..24S.689A} {24, S689}

\bibitem[\protect\citeauthoryear{{Ajith} et~al.,}{{Ajith} et~al.}{2011}]{2011PhRvL.106x1101A}
{Ajith} P.,  et~al., 2011, \mn@doi [\prl] {10.1103/PhysRevLett.106.241101}, \href {https://ui.adsabs.harvard.edu/abs/2011PhRvL.106x1101A} {106, 241101}

\bibitem[\protect\citeauthoryear{{Aky{\"u}z} et~al.,}{{Aky{\"u}z} et~al.}{2025}]{Akyuz2025}
{Aky{\"u}z} A.,  et~al., 2025, \mn@doi [arXiv e-prints] {10.48550/arXiv.2507.08778}, \href {https://ui.adsabs.harvard.edu/abs/2025arXiv250708778A} {p. arXiv:2507.08778}

\bibitem[\protect\citeauthoryear{{Ayzenberg} et~al.,}{{Ayzenberg} et~al.}{2025}]{Ayzenberg2025}
{Ayzenberg} D.,  et~al., 2025, \mn@doi [Living Reviews in Relativity] {10.1007/s41114-025-00057-0}, \href {https://ui.adsabs.harvard.edu/abs/2025LRR....28....4A} {28, 4}

\bibitem[\protect\citeauthoryear{{Barack} et~al.,}{{Barack} et~al.}{2019}]{Barack2019}
{Barack} L.,  et~al., 2019, \mn@doi [Classical and Quantum Gravity] {10.1088/1361-6382/ab0587}, \href {https://ui.adsabs.harvard.edu/abs/2019CQGra..36n3001B} {36, 143001}

\bibitem[\protect\citeauthoryear{Barrault}{Barrault}{2004}]{BARRAULT2004}
Barrault M.,  2004, \mn@doi [Comptes Rendus Mathematique] {10.1016/j.crma.2004.08.006}, 339, 667

\bibitem[\protect\citeauthoryear{{Blackman} et~al.,}{{Blackman} et~al.}{2017}]{Blackman2017}
{Blackman} J.,  et~al., 2017, \mn@doi [\prd] {10.1103/PhysRevD.96.024058}, \href {https://ui.adsabs.harvard.edu/abs/2017PhRvD..96b4058B} {96, 024058}

\bibitem[\protect\citeauthoryear{{Blanchet}}{{Blanchet}}{2006}]{2006LRR.....9....4B}
{Blanchet} L.,  2006, \mn@doi [Living Reviews in Relativity] {10.12942/lrr-2006-4}, \href {https://ui.adsabs.harvard.edu/abs/2006LRR.....9....4B} {9, 4}

\bibitem[\protect\citeauthoryear{{Blanchet}, {Iyer}, {Will}  \& {Wiseman}}{{Blanchet} et~al.}{1996}]{1996CQGra..13..575B}
{Blanchet} L.,  {Iyer} B.~R.,  {Will} C.~M.,   {Wiseman} A.~G.,  1996, \mn@doi [Classical and Quantum Gravity] {10.1088/0264-9381/13/4/002}, \href {https://ui.adsabs.harvard.edu/abs/1996CQGra..13..575B} {13, 575}

\bibitem[\protect\citeauthoryear{{Boh{\'e}} et~al.,}{{Boh{\'e}} et~al.}{2017}]{2017PhRvD..95d4028B}
{Boh{\'e}} A.,  et~al., 2017, \mn@doi [\prd] {10.1103/PhysRevD.95.044028}, \href {https://ui.adsabs.harvard.edu/abs/2017PhRvD..95d4028B} {95, 044028}

\bibitem[\protect\citeauthoryear{{Boyle}, {Buonanno}, {Kidder}, {Mrou{\'e}}, {Pan}, {Pfeiffer}  \& {Scheel}}{{Boyle} et~al.}{2008}]{Boyle2008}
{Boyle} M.,  {Buonanno} A.,  {Kidder} L.~E.,  {Mrou{\'e}} A.~H.,  {Pan} Y.,  {Pfeiffer} H.~P.,   {Scheel} M.~A.,  2008, \mn@doi [\prd] {10.1103/PhysRevD.78.104020}, \href {https://ui.adsabs.harvard.edu/abs/2008PhRvD..78j4020B} {78, 104020}

\bibitem[\protect\citeauthoryear{{Buchman}, {Pfeiffer}, {Scheel}  \& {Szil{\'a}gyi}}{{Buchman} et~al.}{2012}]{2012PhRvD..86h4033B}
{Buchman} L.~T.,  {Pfeiffer} H.~P.,  {Scheel} M.~A.,   {Szil{\'a}gyi} B.,  2012, \mn@doi [\prd] {10.1103/PhysRevD.86.084033}, \href {https://ui.adsabs.harvard.edu/abs/2012PhRvD..86h4033B} {86, 084033}

\bibitem[\protect\citeauthoryear{{Buonanno} \& {Damour}}{{Buonanno} \& {Damour}}{1999}]{1999PhRvD..59h4006B}
{Buonanno} A.,  {Damour} T.,  1999, \mn@doi [\prd] {10.1103/PhysRevD.59.084006}, \href {https://ui.adsabs.harvard.edu/abs/1999PhRvD..59h4006B} {59, 084006}

\bibitem[\protect\citeauthoryear{{Buonanno}, {Cook}  \& {Pretorius}}{{Buonanno} et~al.}{2007}]{Buonanno2007}
{Buonanno} A.,  {Cook} G.~B.,   {Pretorius} F.,  2007, \mn@doi [\prd] {10.1103/PhysRevD.75.124018}, \href {https://ui.adsabs.harvard.edu/abs/2007PhRvD..75l4018B} {75, 124018}

\bibitem[\protect\citeauthoryear{{Da Re} et~al.,}{{Da Re} et~al.}{2025}]{DaRe2025}
{Da Re} G.,  et~al., 2025, \mn@doi [\prd] {10.1103/PhysRevD.111.124019}, \href {https://ui.adsabs.harvard.edu/abs/2025PhRvD.111l4019D} {111, 124019}

\bibitem[\protect\citeauthoryear{{Dayanga} \& {Bose}}{{Dayanga} \& {Bose}}{2010}]{2010APS..NWS.H1012D}
{Dayanga} T.,  {Bose} S.,  2010, in APS Northwest Section Meeting Abstracts. p. H1.012

\bibitem[\protect\citeauthoryear{{Einstein}}{{Einstein}}{1916a}]{1916SPAW.......688E}
{Einstein} A.,  1916a, Sitzungsberichte der K{\"o}niglich Preu{\ss}ischen Akademie der Wissenschaften (Berlin, \href {https://ui.adsabs.harvard.edu/abs/1916SPAW.......688E} {pp 688--696}

\bibitem[\protect\citeauthoryear{{Einstein}}{{Einstein}}{1916b}]{1916AnP...354..769E}
{Einstein} A.,  1916b, \mn@doi [Annalen der Physik] {10.1002/andp.19163540702}, \href {https://ui.adsabs.harvard.edu/abs/1916AnP...354..769E} {354, 769}

\bibitem[\protect\citeauthoryear{{Field}, {Galley}, {Herrmann}, {Hesthaven}, {Ochsner}  \& {Tiglio}}{{Field} et~al.}{2011}]{2011PhRvL.106v1102F}
{Field} S.~E.,  {Galley} C.~R.,  {Herrmann} F.,  {Hesthaven} J.~S.,  {Ochsner} E.,   {Tiglio} M.,  2011, \mn@doi [\prl] {10.1103/PhysRevLett.106.221102}, \href {https://ui.adsabs.harvard.edu/abs/2011PhRvL.106v1102F} {106, 221102}

\bibitem[\protect\citeauthoryear{{Field}, {Galley}, {Hesthaven}, {Kaye}  \& {Tiglio}}{{Field} et~al.}{2014}]{2014PhRvX...4c1006F}
{Field} S.~E.,  {Galley} C.~R.,  {Hesthaven} J.~S.,  {Kaye} J.,   {Tiglio} M.,  2014, \mn@doi [Physical Review X] {10.1103/PhysRevX.4.031006}, \href {https://ui.adsabs.harvard.edu/abs/2014PhRvX...4c1006F} {4, 031006}

\bibitem[\protect\citeauthoryear{GWOSC}{GWOSC}{2019}]{gwosc}
GWOSC 2019, Data release for event GW150914, https://doi.org/10.7935/K5MW2F23, \url {https://doi.org/10.7935/K5MW2F23}

\bibitem[\protect\citeauthoryear{{Gallegos-Garcia}, {Berry}, {Marchant}  \& {Kalogera}}{{Gallegos-Garcia} et~al.}{2021}]{Gallegos2021}
{Gallegos-Garcia} M.,  {Berry} C. P.~L.,  {Marchant} P.,   {Kalogera} V.,  2021, \mn@doi [\apj] {10.3847/1538-4357/ac2610}, \href {https://ui.adsabs.harvard.edu/abs/2021ApJ...922..110G} {922, 110}

\bibitem[\protect\citeauthoryear{{Garc{\'\i}a}, {Simaz Bunzel}, {Chaty}, {Porter}  \& {Chassande-Mottin}}{{Garc{\'\i}a} et~al.}{2021}]{Garcia2021}
{Garc{\'\i}a} F.,  {Simaz Bunzel} A.,  {Chaty} S.,  {Porter} E.,   {Chassande-Mottin} E.,  2021, \mn@doi [\aap] {10.1051/0004-6361/202038357}, \href {https://ui.adsabs.harvard.edu/abs/2021A&A...649A.114G} {649, A114}

\bibitem[\protect\citeauthoryear{{Gerosa} \& {Fishbach}}{{Gerosa} \& {Fishbach}}{2021}]{Gerosa2021}
{Gerosa} D.,  {Fishbach} M.,  2021, \mn@doi [Nature Astronomy] {10.1038/s41550-021-01398-w}, \href {https://ui.adsabs.harvard.edu/abs/2021NatAs...5..749G} {5, 749}

\bibitem[\protect\citeauthoryear{{Giesler}, {Isi}, {Scheel}  \& {Teukolsky}}{{Giesler} et~al.}{2019}]{Giesler2019}
{Giesler} M.,  {Isi} M.,  {Scheel} M.~A.,   {Teukolsky} S.~A.,  2019, \mn@doi [Physical Review X] {10.1103/PhysRevX.9.041060}, \href {https://ui.adsabs.harvard.edu/abs/2019PhRvX...9d1060G} {9, 041060}

\bibitem[\protect\citeauthoryear{{Hannam}, {Schmidt}, {Boh{\'e}}, {Haegel}, {Husa}, {Ohme}, {Pratten}  \& {P{\"u}rrer}}{{Hannam} et~al.}{2014}]{2014PhRvL.113o1101H}
{Hannam} M.,  {Schmidt} P.,  {Boh{\'e}} A.,  {Haegel} L.,  {Husa} S.,  {Ohme} F.,  {Pratten} G.,   {P{\"u}rrer} M.,  2014, \mn@doi [\prl] {10.1103/PhysRevLett.113.151101}, \href {https://ui.adsabs.harvard.edu/abs/2014PhRvL.113o1101H} {113, 151101}

\bibitem[\protect\citeauthoryear{{Healy} \& {Lousto}}{{Healy} \& {Lousto}}{2017}]{Healy2017}
{Healy} J.,  {Lousto} C.~O.,  2017, \mn@doi [\prd] {10.1103/PhysRevD.95.024037}, \href {https://ui.adsabs.harvard.edu/abs/2017PhRvD..95b4037H} {95, 024037}

\bibitem[\protect\citeauthoryear{{Hinder} et~al.,}{{Hinder} et~al.}{2013}]{2013CQGra..31b5012H}
{Hinder} I.,  et~al., 2013, \mn@doi [Classical and Quantum Gravity] {10.1088/0264-9381/31/2/025012}, \href {https://ui.adsabs.harvard.edu/abs/2013CQGra..31b5012H} {31, 025012}

\bibitem[\protect\citeauthoryear{{Jim{\'e}nez-Forteza}, {Keitel}, {Husa}, {Hannam}, {Khan}  \& {P{\"u}rrer}}{{Jim{\'e}nez-Forteza} et~al.}{2017}]{Jimenez2017}
{Jim{\'e}nez-Forteza} X.,  {Keitel} D.,  {Husa} S.,  {Hannam} M.,  {Khan} S.,   {P{\"u}rrer} M.,  2017, \mn@doi [\prd] {10.1103/PhysRevD.95.064024}, \href {https://ui.adsabs.harvard.edu/abs/2017PhRvD..95f4024J} {95, 064024}

\bibitem[\protect\citeauthoryear{{Khan}, {Husa}, {Hannam}, {Ohme}, {P{\"u}rrer}, {Forteza}  \& {Boh{\'e}}}{{Khan} et~al.}{2016}]{2016PhRvD..93d4007K}
{Khan} S.,  {Husa} S.,  {Hannam} M.,  {Ohme} F.,  {P{\"u}rrer} M.,  {Forteza} X.~J.,   {Boh{\'e}} A.,  2016, \mn@doi [\prd] {10.1103/PhysRevD.93.044007}, \href {https://ui.adsabs.harvard.edu/abs/2016PhRvD..93d4007K} {93, 044007}

\bibitem[\protect\citeauthoryear{{Klencki}, {Podsiadlowski}, {Langer}, {Olejak}, {Justham}, {Vigna-G{\'o}mez}  \& {de Mink}}{{Klencki} et~al.}{2025}]{Klencki2025}
{Klencki} J.,  {Podsiadlowski} P.,  {Langer} N.,  {Olejak} A.,  {Justham} S.,  {Vigna-G{\'o}mez} A.,   {de Mink} S.~E.,  2025, \mn@doi [arXiv e-prints] {10.48550/arXiv.2505.08860}, \href {https://ui.adsabs.harvard.edu/abs/2025arXiv250508860K} {p. arXiv:2505.08860}

\bibitem[\protect\citeauthoryear{{Kumar}, {Barkett}, {Bhagwat}, {Afshari}, {Brown}, {Lovelace}, {Scheel}  \& {Szil{\'a}gyi}}{{Kumar} et~al.}{2015}]{2015PhRvD..92j2001K}
{Kumar} P.,  {Barkett} K.,  {Bhagwat} S.,  {Afshari} N.,  {Brown} D.~A.,  {Lovelace} G.,  {Scheel} M.~A.,   {Szil{\'a}gyi} B.,  2015, \mn@doi [\prd] {10.1103/PhysRevD.92.102001}, \href {https://ui.adsabs.harvard.edu/abs/2015PhRvD..92j2001K} {92, 102001}

\bibitem[\protect\citeauthoryear{{Li}, {Lin}  \& {Yuan}}{{Li} et~al.}{2023}]{Li2023}
{Li} G.-P.,  {Lin} D.-B.,   {Yuan} Y.,  2023, \mn@doi [\prd] {10.1103/PhysRevD.107.063007}, \href {https://ui.adsabs.harvard.edu/abs/2023PhRvD.107f3007L} {107, 063007}

\bibitem[\protect\citeauthoryear{{Littenberg}, {Baker}, {Buonanno}  \& {Kelly}}{{Littenberg} et~al.}{2013}]{2013PhRvD..87j4003L}
{Littenberg} T.~B.,  {Baker} J.~G.,  {Buonanno} A.,   {Kelly} B.~J.,  2013, \mn@doi [\prd] {10.1103/PhysRevD.87.104003}, \href {https://ui.adsabs.harvard.edu/abs/2013PhRvD..87j4003L} {87, 104003}

\bibitem[\protect\citeauthoryear{{Lovelace}, {Boyle}, {Scheel}  \& {Szil{\'a}gyi}}{{Lovelace} et~al.}{2012}]{2012CQGra..29d5003L}
{Lovelace} G.,  {Boyle} M.,  {Scheel} M.~A.,   {Szil{\'a}gyi} B.,  2012, \mn@doi [Classical and Quantum Gravity] {10.1088/0264-9381/29/4/045003}, \href {https://ui.adsabs.harvard.edu/abs/2012CQGra..29d5003L} {29, 045003}

\bibitem[\protect\citeauthoryear{{McKernan}, {Ford}, {Callister}, {Farr}, {O'Shaughnessy}, {Smith}, {Thrane}  \& {Vajpeyi}}{{McKernan} et~al.}{2022}]{McKernan2022}
{McKernan} B.,  {Ford} K.~E.~S.,  {Callister} T.,  {Farr} W.~M.,  {O'Shaughnessy} R.,  {Smith} R.,  {Thrane} E.,   {Vajpeyi} A.,  2022, \mn@doi [\mnras] {10.1093/mnras/stac1570}, \href {https://ui.adsabs.harvard.edu/abs/2022MNRAS.514.3886M} {514, 3886}

\bibitem[\protect\citeauthoryear{{Mitman} et~al.,}{{Mitman} et~al.}{2021}]{Mitman2021}
{Mitman} K.,  et~al., 2021, \mn@doi [\prd] {10.1103/PhysRevD.103.024031}, \href {https://ui.adsabs.harvard.edu/abs/2021PhRvD.103b4031M} {103, 024031}

\bibitem[\protect\citeauthoryear{{Mukherjee}, {Phukon}, {Datta}  \& {Bose}}{{Mukherjee} et~al.}{2024}]{Mukherjee2024}
{Mukherjee} S.,  {Phukon} K.~S.,  {Datta} S.,   {Bose} S.,  2024, \mn@doi [\prd] {10.1103/PhysRevD.110.124027}, \href {https://ui.adsabs.harvard.edu/abs/2024PhRvD.110l4027M} {110, 124027}

\bibitem[\protect\citeauthoryear{Nitz et~al.,}{Nitz et~al.}{2019}]{alex_nitz_2019_3265452}
Nitz A.,  et~al., 2019, gwastro/pycbc: PyCBC Release v1.14.1, \mn@doi{10.5281/zenodo.3265452}, \url {https://doi.org/10.5281/zenodo.3265452}

\bibitem[\protect\citeauthoryear{{Ossokine}, {Boyle}, {Kidder}, {Pfeiffer}, {Scheel}  \& {Szil{\'a}gyi}}{{Ossokine} et~al.}{2015}]{2015PhRvD..92j4028O}
{Ossokine} S.,  {Boyle} M.,  {Kidder} L.~E.,  {Pfeiffer} H.~P.,  {Scheel} M.~A.,   {Szil{\'a}gyi} B.,  2015, \mn@doi [\prd] {10.1103/PhysRevD.92.104028}, \href {https://ui.adsabs.harvard.edu/abs/2015PhRvD..92j4028O} {92, 104028}

\bibitem[\protect\citeauthoryear{{Owen} \& {Sathyaprakash}}{{Owen} \& {Sathyaprakash}}{1999}]{1999PhRvD..60b2002O}
{Owen} B.~J.,  {Sathyaprakash} B.~S.,  1999, \mn@doi [\prd] {10.1103/PhysRevD.60.022002}, \href {https://ui.adsabs.harvard.edu/abs/1999PhRvD..60b2002O} {60, 022002}

\bibitem[\protect\citeauthoryear{{Pan}, {Buonanno}, {Boyle}, {Buchman}, {Kidder}, {Pfeiffer}  \& {Scheel}}{{Pan} et~al.}{2011}]{2011PhRvD..84l4052P}
{Pan} Y.,  {Buonanno} A.,  {Boyle} M.,  {Buchman} L.~T.,  {Kidder} L.~E.,  {Pfeiffer} H.~P.,   {Scheel} M.~A.,  2011, \mn@doi [\prd] {10.1103/PhysRevD.84.124052}, \href {https://ui.adsabs.harvard.edu/abs/2011PhRvD..84l4052P} {84, 124052}

\bibitem[\protect\citeauthoryear{{Pan}, {Buonanno}, {Taracchini}, {Kidder}, {Mrou{\'e}}, {Pfeiffer}, {Scheel}  \& {Szil{\'a}gyi}}{{Pan} et~al.}{2014}]{2014PhRvD..89h4006P}
{Pan} Y.,  {Buonanno} A.,  {Taracchini} A.,  {Kidder} L.~E.,  {Mrou{\'e}} A.~H.,  {Pfeiffer} H.~P.,  {Scheel} M.~A.,   {Szil{\'a}gyi} B.,  2014, \mn@doi [\prd] {10.1103/PhysRevD.89.084006}, \href {https://ui.adsabs.harvard.edu/abs/2014PhRvD..89h4006P} {89, 084006}

\bibitem[\protect\citeauthoryear{Poisson \& Will}{Poisson \& Will}{2014}]{poisson2014gravity}
Poisson E.,  Will C.,  2014, Gravity: Newtonian, Post-Newtonian, Relativistic.
Cambridge University Press, \url {https://books.google.com.tr/books?id=PZ5cAwAAQBAJ}

\bibitem[\protect\citeauthoryear{{Pretorius}}{{Pretorius}}{2005}]{2005PhRvL..95l1101P}
{Pretorius} F.,  2005, \mn@doi [\prl] {10.1103/PhysRevLett.95.121101}, \href {https://ui.adsabs.harvard.edu/abs/2005PhRvL..95l1101P} {95, 121101}

\bibitem[\protect\citeauthoryear{Prud'Homme, Rovas, Veroy, Machiels, Maday, Patera  \& Turinici}{Prud'Homme et~al.}{2001}]{prudhomme:hal-00798326}
Prud'Homme C.,  Rovas D.~V.,  Veroy K.,  Machiels L.,  Maday Y.,  Patera A.~T.,   Turinici G.,  2001, \mn@doi [{Journal of Fluids Engineering}] {10.1115/1.1448332}, 124, 70

\bibitem[\protect\citeauthoryear{PyCBC}{PyCBC}{2021b}]{bandpass}
PyCBC 2021b, Applying highpass / lowpass filters, https://pycbc.org/pycbc/latest/html/filter.html, \url {https://pycbc.org/pycbc/latest/html/filter.html}

\bibitem[\protect\citeauthoryear{PyCBC}{PyCBC}{2021c}]{signal_tutorials}
PyCBC 2021c, Generating Waveforms and Matched Filtering, https://github.com/gwastro/PyCBC-Tutorials, \url {https://github.com/gwastro/PyCBC-Tutorials}

\bibitem[\protect\citeauthoryear{PyCBC}{PyCBC}{2021d}]{gpstime}
PyCBC 2021d, GWOSC, UTC/GPS Time Converter, https://www.gw-openscience.org/gps/, \url {https://www.gw-openscience.org/gps/}

\bibitem[\protect\citeauthoryear{PyCBC}{PyCBC}{2021a}]{smoothed_strain}
PyCBC 2021a, Signal Processing with GW150914, https://pycbc.org/pycbc/latest/html/gw150914.html, \url {https://pycbc.org/pycbc/latest/html/gw150914.html}

\bibitem[\protect\citeauthoryear{{Raymond}, {van der Sluys}, {Mandel}, {Kalogera}, {R{\"o}ver}  \& {Christensen}}{{Raymond} et~al.}{2009}]{2009CQGra..26k4007R}
{Raymond} V.,  {van der Sluys} M.~V.,  {Mandel} I.,  {Kalogera} V.,  {R{\"o}ver} C.,   {Christensen} N.,  2009, \mn@doi [Classical and Quantum Gravity] {10.1088/0264-9381/26/11/114007}, \href {https://ui.adsabs.harvard.edu/abs/2009CQGra..26k4007R} {26, 114007}

\bibitem[\protect\citeauthoryear{{Scheel}, {Boyle}, {Chu}, {Kidder}, {Matthews}  \& {Pfeiffer}}{{Scheel} et~al.}{2009}]{2009PhRvD..79b4003S}
{Scheel} M.~A.,  {Boyle} M.,  {Chu} T.,  {Kidder} L.~E.,  {Matthews} K.~D.,   {Pfeiffer} H.~P.,  2009, \mn@doi [\prd] {10.1103/PhysRevD.79.024003}, \href {https://ui.adsabs.harvard.edu/abs/2009PhRvD..79b4003S} {79, 024003}

\bibitem[\protect\citeauthoryear{{Smith}, {Cannon}, {Hanna}, {Keppel}  \& {Mandel}}{{Smith} et~al.}{2013}]{2013PhRvD..87l2002S}
{Smith} R.~J.~E.,  {Cannon} K.,  {Hanna} C.,  {Keppel} D.,   {Mandel} I.,  2013, \mn@doi [\prd] {10.1103/PhysRevD.87.122002}, \href {https://ui.adsabs.harvard.edu/abs/2013PhRvD..87l2002S} {87, 122002}

\bibitem[\protect\citeauthoryear{{Taracchini} et~al.,}{{Taracchini} et~al.}{2012}]{2012PhRvD..86b4011T}
{Taracchini} A.,  et~al., 2012, \mn@doi [\prd] {10.1103/PhysRevD.86.024011}, \href {https://ui.adsabs.harvard.edu/abs/2012PhRvD..86b4011T} {86, 024011}

\bibitem[\protect\citeauthoryear{{Taracchini} et~al.,}{{Taracchini} et~al.}{2013}]{2013APS..APRC10004T}
{Taracchini} A.,  et~al., 2013, in APS April Meeting Abstracts. p. C10.004

\bibitem[\protect\citeauthoryear{{Taracchini} et~al.,}{{Taracchini} et~al.}{2014}]{2014PhRvD..89f1502T}
{Taracchini} A.,  et~al., 2014, \mn@doi [\prd] {10.1103/PhysRevD.89.061502}, \href {https://ui.adsabs.harvard.edu/abs/2014PhRvD..89f1502T} {89, 061502}

\bibitem[\protect\citeauthoryear{{The LIGO Scientific Collaboration} et~al.,}{{The LIGO Scientific Collaboration} et~al.}{2019}]{2019arXiv191211716T}
{The LIGO Scientific Collaboration} et~al., 2019, arXiv e-prints, \href {https://ui.adsabs.harvard.edu/abs/2019arXiv191211716T} {p. arXiv:1912.11716}

\bibitem[\protect\citeauthoryear{{Varma}, {Field}, {Scheel}, {Blackman}, {Gerosa}, {Stein}, {Kidder}  \& {Pfeiffer}}{{Varma} et~al.}{2019}]{Varma2019}
{Varma} V.,  {Field} S.~E.,  {Scheel} M.~A.,  {Blackman} J.,  {Gerosa} D.,  {Stein} L.~C.,  {Kidder} L.~E.,   {Pfeiffer} H.~P.,  2019, \mn@doi [Physical Review Research] {10.1103/PhysRevResearch.1.033015}, \href {https://ui.adsabs.harvard.edu/abs/2019PhRvR...1c3015V} {1, 033015}

\bibitem[\protect\citeauthoryear{{Varma}, {Mould}, {Gerosa}, {Scheel}, {Kidder}  \& {Pfeiffer}}{{Varma} et~al.}{2021}]{Varma2021}
{Varma} V.,  {Mould} M.,  {Gerosa} D.,  {Scheel} M.~A.,  {Kidder} L.~E.,   {Pfeiffer} H.~P.,  2021, \mn@doi [\prd] {10.1103/PhysRevD.103.064003}, \href {https://ui.adsabs.harvard.edu/abs/2021PhRvD.103f4003V} {103, 064003}

\bibitem[\protect\citeauthoryear{Welch}{Welch}{1967}]{1161901}
Welch P.,  1967, \mn@doi [IEEE Transactions on Audio and Electroacoustics] {10.1109/TAU.1967.1161901}, 15, 70

\bibitem[\protect\citeauthoryear{{Yakut}}{{Yakut}}{2010}]{Yakut2010}
{Yakut} K.,  2010, in {Kalogera} V.,  {van der Sluys} M.,  eds,  American Institute of Physics Conference Series Vol. 1314, International Conference on Binaries: in celebration of Ron Webbink's 65th Birthday. AIP, pp 299--300, \mn@doi{10.1063/1.3536389}

\bibitem[\protect\citeauthoryear{{Yakut}, {Kalomeni}  \& {Tout}}{{Yakut} et~al.}{2008}]{Yakut2008}
{Yakut} K.,  {Kalomeni} B.,   {Tout} C.~A.,  2008, \mn@doi [arXiv e-prints] {10.48550/arXiv.0811.0455}, \href {https://ui.adsabs.harvard.edu/abs/2008arXiv0811.0455Y} {p. arXiv:0811.0455}

\bibitem[\protect\citeauthoryear{{{\c{C}}okluk}, {Yakut}  \& {Giacomazzo}}{{{\c{C}}okluk} et~al.}{2024}]{Cokluk2024}
{{\c{C}}okluk} K.~A.,  {Yakut} K.,   {Giacomazzo} B.,  2024, \mn@doi [\mnras] {10.1093/mnras/stad3752}, \href {https://ui.adsabs.harvard.edu/abs/2024MNRAS.527.8043C} {527, 8043}

\bibitem[\protect\citeauthoryear{{van der Sluys}, {Mandel}, {Raymond}, {Kalogera}, {R{\"o}ver}  \& {Christensen}}{{van der Sluys} et~al.}{2009}]{2009CQGra..26t4010V}
{van der Sluys} M.,  {Mandel} I.,  {Raymond} V.,  {Kalogera} V.,  {R{\"o}ver} C.,   {Christensen} N.,  2009, \mn@doi [Classical and Quantum Gravity] {10.1088/0264-9381/26/20/204010}, \href {https://ui.adsabs.harvard.edu/abs/2009CQGra..26t4010V} {26, 204010}

\makeatother
\end{thebibliography}

\section*{Supplementary}
\label{sec:onlinefigures}
\begin{table*}
	\renewcommand{\arraystretch}{0.88}
	\setlength{\tabcolsep}{0.050in}
	\small
	\caption{Here, h$_{\rm{max}}$ is the maximum wave amplitude, in order of $\times 10^{-18}$. Also $\chi_{1i}$ is the initial spin parameter of the massive BH component of the system.  M$_{\rm{tot}}/M_\odot$ is the initial total mass. M$_{\rm{FL}}$ is the fractional mass loss and $\chi_{\rm{f}}$ is the final spin parameter. The table is obtained from the modeled data according to $PN$ case.} 
	\hspace{-1.2cm}
	\begin{tabular}{llllllllllllll}
		$\boldsymbol{\chi}_{1i}$ $\downarrow$ &\textbf{M$_{\rm{tot}}$/M$_\odot$}$\downarrow$  &$\textbf{q}\rightarrow$  &\textbf{1.00} &\textbf{1.10} &\textbf{1.20} &\textbf{1.30} &\textbf{1.40} &\textbf{1.50} &\textbf{1.60} &\textbf{1.70} &\textbf{1.80} &\textbf{1.90} &\textbf{2.00}   \\
		\hline
		\textbf{0.00} & ~~~~\textbf{15} & \textbf{h}$_{\rm{max}}$         &0.179 &0.178 &0.177 &0.175 &0.173 &0.170 &0.168 &0.164 &0.161 &0.159 &0.156 \\
		&                 & \textbf{M}$_{\rm{FL}}$ 	     &4.821 &4.803 &4.754 &4.683 &4.596 &4.498 &4.394 &4.285 &4.174 &4.062 &3.951 \\
		&                 & $\boldsymbol{\chi}_{\rm{f}}$   &0.686 &0.685 &0.682 &0.677 &0.671 &0.664 &0.656 &0.648 &0.640 &0.632 &0.623 \\ \cline{2-14}
		& ~~~~\textbf{30} & \textbf{h}$_{\rm{max}}$         &0.357 &0.356 &0.354 &0.350 &0.346 &0.340 &0.335 &0.330 &0.324 &0.318 &0.312 \\
		&                 & \textbf{M}$_{\rm{FL}}$ 	     &4.821 &4.803 &4.754 &4.683 &4.596 &4.498 &4.394 &4.285 &4.174 &4.062 &3.951 \\
		&                 & $\boldsymbol{\chi}_{\rm{f}}$   &0.686 &0.685 &0.682 &0.677 &0.671 &0.664 &0.656 &0.648 &0.640 &0.632 &0.623 \\ \cline{2-14}
		& ~~~~\textbf{60} & \textbf{h}$_{\rm{max}}$         &0.714 &0.712 &0.707 &0.700 &0.691 &0.681 &0.670 &0.659 &0.647 &0.636 &0.624 \\
		&                 & \textbf{M}$_{\rm{FL}}$          &4.821 &4.803 &4.754 &4.683 &4.596 &4.498 &4.394 &4.285 &4.174 &4.062 &3.951 \\
		&                 & $\boldsymbol{\chi}_{\rm{f}}$   &0.686 &0.685 &0.682 &0.677 &0.671 &0.664 &0.656 &0.648 &0.640 &0.632 &0.623 \\ \cline{2-14}
		& ~~~~\textbf{90} & \textbf{h}$_{\rm{max}}$         &1.071 &1.068 &1.061 &1.050 &1.037 &1.022 &1.006 &0.989 &0.971 &0.954 &0.936 \\
		&                 & \textbf{M}$_{\rm{FL}}$ 	     &4.821 &4.803 &4.754 &4.683 &4.596 &4.498 &4.394 &4.285 &4.174 &4.062 &3.951 \\
		&                 & $\boldsymbol{\chi}_{\rm{f}}$   &0.686 &0.685 &0.682 &0.677 &0.671 &0.664 &0.656 &0.648 &0.640 &0.632 &0.623 \\ 
		\hline	
		\textbf{-0.33} & ~~~~\textbf{15} & \textbf{h}$_{\rm{max}}$         &0.179 &0.178 &0.177 &0.175 &0.172 &0.170 &0.168 &0.165 &0.162 &0.159 &0.156 \\
		&                 & \textbf{M}$_{\rm{FL}}$ 	     &4.821 &4.720 &4.600 &4.471 &4.337 &4.202 &4.067 &3.935 &3.807 &3.683 &3.563 \\
		&                 & $\boldsymbol{\chi}_{\rm{f}}$   &0.686 &0.673 &0.659 &0.644 &0.629 &0.613 &0.598 &0.582 &0.567 &0.552 &0.537 \\ \cline{2-14}
		& ~~~~\textbf{30} & \textbf{h}$_{\rm{max}}$         &0.357 &0.356 &0.354 &0.350 &0.346 &0.341 &0.335 &0.330 &0.324 &0.318 &0.312 \\
		&                 & \textbf{M}$_{\rm{FL}}$ 	     &4.821 &4.720 &4.600 &4.471 &4.337 &4.202 &4.067 &3.935 &3.807 &3.683 &3.563 \\
		&                 & $\boldsymbol{\chi}_{\rm{f}}$   &0.686 &0.673 &0.659 &0.644 &0.629 &0.613 &0.598 &0.582 &0.567 &0.552 &0.537 \\ \cline{2-14}
		& ~~~~\textbf{60} & \textbf{h}$_{\rm{max}}$         &0.714 &0.712 &0.707 &0.700 &0.692 &0.682 &0.671 &0.660 &0.648 &0.636 &0.624 \\
		&                 & \textbf{M}$_{\rm{FL}}$          &4.821 &4.720 &4.600 &4.471 &4.337 &4.202 &4.067 &3.935 &3.807 &3.683 &3.563 \\
		&                 & $\boldsymbol{\chi}_{\rm{f}}$   &0.686 &0.673 &0.659 &0.644 &0.629 &0.613 &0.598 &0.582 &0.567 &0.552 &0.537 \\ \cline{2-14}
		& ~~~~\textbf{90} & \textbf{h}$_{\rm{max}}$         &1.071 &1.069 &1.061 &1.051 &1.038 &1.023 &1.006 &0.989 &0.972 &0.954 &0.936 \\
		&                 & \textbf{M}$_{\rm{FL}}$ 	     &4.821 &4.720 &4.600 &4.471 &4.337 &4.202 &4.067 &3.935 &3.807 &3.683 &3.563 \\
		&                 & $\boldsymbol{\chi}_{\rm{f}}$   &0.686 &0.673 &0.659 &0.644 &0.629 &0.613 &0.598 &0.582 &0.567 &0.552 &0.537 \\ 
		\hline	
		\textbf{-0.66} & ~~~~\textbf{15} & \textbf{h}$_{\rm{max}}$         &0.178 &0.178 &0.177 &0.175 &0.173 &0.171 &0.168 &0.165 &0.162 &0.159 &0.156 \\
		&                 & \textbf{M}$_{\rm{FL}}$ 	     &4.821 &4.640 &4.457 &4.279 &4.107 &3.944 &3.789 &3.642 &3.503 &3.373 &3.249 \\
		&                 & $\boldsymbol{\chi}_{\rm{f}}$   &0.686 &0.662 &0.637 &0.612 &0.587 &0.563 &0.539 &0.516 &0.493 &0.471 &0.449 \\ \cline{2-14}
		& ~~~~\textbf{30} & \textbf{h}$_{\rm{max}}$         &0.357 &0.356 &0.354 &0.350 &0.346 &0.341 &0.335 &0.330 &0.324 &0.318 &0.312 \\
		&                 & \textbf{M}$_{\rm{FL}}$ 	     &4.821 &4.640 &4.457 &4.279 &4.107 &3.944 &3.789 &3.642 &3.503 &3.373 &3.249 \\
		&                 & $\boldsymbol{\chi}_{\rm{f}}$   &0.686 &0.662 &0.637 &0.612 &0.587 &0.563 &0.539 &0.516 &0.493 &0.471 &0.449 \\ \cline{2-14}
		& ~~~~\textbf{60} & \textbf{h}$_{\rm{max}}$         &0.714 &0.713 &0.708 &0.701 &0.692 &0.682 &0.671 &0.660 &0.648 &0.636 &0.624 \\
		&                 & \textbf{M}$_{\rm{FL}}$          &4.821 &4.640 &4.457 &4.279 &4.107 &3.944 &3.789 &3.642 &3.503 &3.373 &3.249 \\
		&                 & $\boldsymbol{\chi}_{\rm{f}}$   &0.686 &0.662 &0.637 &0.612 &0.587 &0.563 &0.539 &0.516 &0.493 &0.471 &0.449 \\ \cline{2-14}
		& ~~~~\textbf{90} & \textbf{h}$_{\rm{max}}$         &1.071 &1.069 &1.062 &1.051 &1.038 &1.023 &1.007 &0.990 &0.972 &0.954 &0.936 \\
		&                 & \textbf{M}$_{\rm{FL}}$ 	     &4.821 &4.640 &4.457 &4.279 &4.107 &3.944 &3.789 &3.642 &3.503 &3.373 &3.249 \\
		&                 & $\boldsymbol{\chi}_{\rm{f}}$   &0.686 &0.662 &0.637 &0.612 &0.587 &0.563 &0.539 &0.516 &0.493 &0.471 &0.449 \\ 
		\hline
		\textbf{-0.83} & ~~~~\textbf{15} & \textbf{h}$_{\rm{max}}$         &0.178 &0.178 &0.177 &0.175 &0.173 &0.171 &0.168 &0.165 &0.162 &0.159 &0.156 \\
		&                 & \textbf{M}$_{\rm{FL}}$ 	     &4.821 &4.601 &4.389 &4.189 &4.002 &3.827 &3.664 &3.512 &3.370 &3.237 &3.113 \\
		&                 & $\boldsymbol{\chi}_{\rm{f}}$   &0.686 &0.656 &0.625 &0.595 &0.566 &0.537 &0.509 &0.482 &0.455 &0.430 &0.405 \\ \cline{2-14}
		& ~~~~\textbf{30} & \textbf{h}$_{\rm{max}}$         &0.357 &0.356 &0.354 &0.350 &0.346 &0.341 &0.336 &0.330 &0.324 &0.318 &0.312 \\
		&                 & \textbf{M}$_{\rm{FL}}$ 	     &4.821 &4.601 &4.389 &4.189 &4.002 &3.827 &3.664 &3.512 &3.370 &3.237 &3.113 \\
		&                 & $\boldsymbol{\chi}_{\rm{f}}$   &0.686 &0.656 &0.625 &0.595 &0.566 &0.537 &0.509 &0.482 &0.455 &0.430 &0.405 \\ \cline{2-14}
		& ~~~~\textbf{60} & \textbf{h}$_{\rm{max}}$         &0.714 &0.713 &0.708 &0.701 &0.692 &0.682 &0.671 &0.660 &0.648 &0.636 &0.624 \\
		&                 & \textbf{M}$_{\rm{FL}}$          &4.821 &4.601 &4.389 &4.189 &4.002 &3.827 &3.664 &3.512 &3.370 &3.237 &3.113 \\
		&                 & $\boldsymbol{\chi}_{\rm{f}}$   &0.686 &0.656 &0.625 &0.595 &0.566 &0.537 &0.509 &0.482 &0.455 &0.430 &0.405 \\ \cline{2-14}
		& ~~~~\textbf{90} & \textbf{h}$_{\rm{max}}$         &1.071 &1.069 &1.062 &1.052 &1.039 &1.024 &1.007 &0.990 &0.972 &0.954 &0.936 \\
		&                 & \textbf{M}$_{\rm{FL}}$ 	     &4.821 &4.601 &4.389 &4.189 &4.002 &3.827 &3.664 &3.512 &3.370 &3.237 &3.113 \\
		&                 & $\boldsymbol{\chi}_{\rm{f}}$   &0.686 &0.656 &0.625 &0.595 &0.566 &0.537 &0.509 &0.482 &0.455 &0.430 &0.405 \\ 
		\hline
	\end{tabular}\label{table:PN}
\end{table*}

\begin{table*}
	\renewcommand{\arraystretch}{0.88}
	\setlength{\tabcolsep}{0.050in}
\small
	\caption{Here, h$_{\rm{max}}$ is the maximum wave amplitude, in order of $\times 10^{-18}$. Also $\chi_{1i}$ is the initial spin parameter of the massive BH component of the system.  M$_{\rm{tot}}/M_\odot$ is the initial total mass. M$_{\rm{FL}}$ is the fractional mass loss and $\chi_{\rm{f}}$ is the final spin parameter. The table is obtained from the modeled data according to $BN$ case.}
	\hspace{-1.2cm}
	\begin{tabular}{llllllllllllll}
		$\boldsymbol{\chi}_{1i}$ $\downarrow$ &\textbf{M$_{\rm{tot}}$/M$_\odot$}$\downarrow$  &$\textbf{q}\rightarrow$  &\textbf{1.00} &\textbf{1.10} &\textbf{1.20} &\textbf{1.30} &\textbf{1.40} &\textbf{1.50} &\textbf{1.60} &\textbf{1.70} &\textbf{1.80} &\textbf{1.90} &\textbf{2.00}   \\
		\hline
		\textbf{0.00} & ~~~~\textbf{15} & \textbf{h}$_{\rm{max}}$         &0.179 &0.178 &0.177 &0.175 &0.173 &0.170 &0.168 &0.164 &0.161 &0.159 &0.156 \\
		&                 & \textbf{M}$_{\rm{FL}}$ 	     &4.821 &4.803 &4.754 &4.683 &4.596 &4.498 &4.394 &4.285 &4.174 &4.062 &3.951 \\
		&                 & $\boldsymbol{\chi}_{\rm{f}}$   &0.686 &0.685 &0.682 &0.677 &0.671 &0.664 &0.656 &0.648 &0.640 &0.632 &0.623 \\ \cline{2-14}
		& ~~~~\textbf{30} & \textbf{h}$_{\rm{max}}$         &0.357 &0.356 &0.354 &0.350 &0.346 &0.340 &0.335 &0.330 &0.324 &0.318 &0.312 \\
		&                 & \textbf{M}$_{\rm{FL}}$ 	     &4.821 &4.803 &4.754 &4.683 &4.596 &4.498 &4.394 &4.285 &4.174 &4.062 &3.951 \\
		&                 & $\boldsymbol{\chi}_{\rm{f}}$   &0.686 &0.685 &0.682 &0.677 &0.671 &0.664 &0.656 &0.648 &0.640 &0.632 &0.623 \\ \cline{2-14}
		& ~~~~\textbf{60} & \textbf{h}$_{\rm{max}}$         &0.714 &0.712 &0.707 &0.700 &0.691 &0.681 &0.670 &0.659 &0.647 &0.636 &0.624 \\
		&                 & \textbf{M}$_{\rm{FL}}$          &4.821 &4.803 &4.754 &4.683 &4.596 &4.498 &4.394 &4.285 &4.174 &4.062 &3.951 \\
		&                 & $\boldsymbol{\chi}_{\rm{f}}$   &0.686 &0.685 &0.682 &0.677 &0.671 &0.664 &0.656 &0.648 &0.640 &0.632 &0.623 \\ \cline{2-14}
		& ~~~~\textbf{90} & \textbf{h}$_{\rm{max}}$         &1.071 &1.068 &1.061 &1.050 &1.037 &1.022 &1.006 &0.989 &0.971 &0.954 &0.936 \\
		&                 & \textbf{M}$_{\rm{FL}}$ 	     &4.821 &4.803 &4.754 &4.683 &4.596 &4.498 &4.394 &4.285 &4.174 &4.062 &3.951 \\
		&                 & $\boldsymbol{\chi}_{\rm{f}}$   &0.686 &0.685 &0.682 &0.677 &0.671 &0.664 &0.656 &0.648 &0.640 &0.632 &0.623 \\ 
		\hline	
		\textbf{-0.33} & ~~~~\textbf{15} & \textbf{h}$_{\rm{max}}$         &0.179 &0.178 &0.177 &0.175 &0.173 &0.171 &0.168 &0.165 &0.162 &0.159 &0.156 \\
		&                 & \textbf{M}$_{\rm{FL}}$ 	     &4.074 &4.058 &4.017 &3.958 &3.886 &3.804 &3.717 &3.626 &3.533 &3.440 &3.347 \\
		&                 & $\boldsymbol{\chi}_{\rm{f}}$   &0.582 &0.581 &0.576 &0.569 &0.560 &0.551 &0.540 &0.529 &0.518 &0.506 &0.494 \\ \cline{2-14}
		& ~~~~\textbf{30} & \textbf{h}$_{\rm{max}}$         &0.358 &0.357 &0.354 &0.351 &0.346 &0.341 &0.336 &0.330 &0.324 &0.318 &0.312 \\
		&                 & \textbf{M}$_{\rm{FL}}$ 	     &4.074 &4.058 &4.017 &3.958 &3.886 &3.804 &3.717 &3.626 &3.533 &3.440 &3.347 \\
		&                 & $\boldsymbol{\chi}_{\rm{f}}$   &0.582 &0.581 &0.576 &0.569 &0.560 &0.551 &0.540 &0.529 &0.518 &0.506 &0.494 \\ \cline{2-14}
		& ~~~~\textbf{60} & \textbf{h}$_{\rm{max}}$         &0.716 &0.714 &0.709 &0.702 &0.693 &0.682 &0.671 &0.660 &0.648 &0.636 &0.624 \\
		&                 & \textbf{M}$_{\rm{FL}}$          &4.074 &4.058 &4.017 &3.958 &3.886 &3.804 &3.717 &3.626 &3.533 &3.440 &3.347 \\
		&                 & $\boldsymbol{\chi}_{\rm{f}}$   &0.582 &0.581 &0.576 &0.569 &0.560 &0.551 &0.540 &0.529 &0.518 &0.506 &0.494 \\ \cline{2-14}
		& ~~~~\textbf{90} & \textbf{h}$_{\rm{max}}$         &1.074 &1.071 &1.064 &1.053 &1.039 &1.024 &1.007 &0.990 &0.972 &0.954 &0.936 \\
		&                 & \textbf{M}$_{\rm{FL}}$ 	     &4.074 &4.058 &4.017 &3.958 &3.886 &3.804 &3.717 &3.626 &3.533 &3.440 &3.347 \\
		&                 & $\boldsymbol{\chi}_{\rm{f}}$   &0.582 &0.581 &0.576 &0.569 &0.560 &0.551 &0.540 &0.529 &0.518 &0.506 &0.494 \\ 
		\hline
		\textbf{-0.66} & ~~~~\textbf{15} & \textbf{h}$_{\rm{max}}$         &0.179 &0.179 &0.177 &0.176 &0.173 &0.171 &0.168 &0.165 &0.162 &0.159 &0.156 \\
		&                 & \textbf{M}$_{\rm{FL}}$ 	     &3.537 &3.523 &3.488 &3.437 &3.375 &3.306 &3.231 &3.152 &3.073 &2.992 &2.912 \\
		&                 & $\boldsymbol{\chi}_{\rm{f}}$   &0.473 &0.471 &0.465 &0.456 &0.445 &0.433 &0.419 &0.405 &0.391 &0.376 &0.361 \\ \cline{2-14}
		& ~~~~\textbf{30} & \textbf{h}$_{\rm{max}}$         &0.359 &0.358 &0.355 &0.351 &0.347 &0.341 &0.336 &0.330 &0.324 &0.318 &0.311 \\
		&                 & \textbf{M}$_{\rm{FL}}$ 	     &3.537 &3.523 &3.488 &3.437 &3.375 &3.306 &3.231 &3.152 &3.072 &2.992 &2.912 \\
		&                 & $\boldsymbol{\chi}_{\rm{f}}$   &0.473 &0.471 &0.465 &0.456 &0.445 &0.433 &0.419 &0.405 &0.391 &0.376 &0.361 \\ \cline{2-14}
		& ~~~~\textbf{60} & \textbf{h}$_{\rm{max}}$         &0.718 &0.716 &0.711 &0.703 &0.694 &0.683 &0.672 &0.660 &0.648 &0.635 &0.623 \\
		&                 & \textbf{M}$_{\rm{FL}}$          &3.537 &3.523 &3.488 &3.437 &3.375 &3.306 &3.231 &3.152 &3.072 &2.992 &2.912 \\
		&                 & $\boldsymbol{\chi}_{\rm{f}}$   &0.473 &0.471 &0.465 &0.456 &0.445 &0.433 &0.419 &0.405 &0.391 &0.376 &0.361 \\ \cline{2-14}
		& ~~~~\textbf{90} & \textbf{h}$_{\rm{max}}$         &1.077 &1.074 &1.066 &1.055 &1.041 &1.025 &1.008 &0.990 &0.972 &0.953 &0.935 \\
		&                 & \textbf{M}$_{\rm{FL}}$ 	     &3.537 &3.523 &3.488 &3.437 &3.375 &3.306 &3.231 &3.152 &3.072 &2.992 &2.912 \\
		&                 & $\boldsymbol{\chi}_{\rm{f}}$   &0.473 &0.471 &0.465 &0.456 &0.445 &0.433 &0.419 &0.405 &0.391 &0.376 &0.361 \\ 
		\hline
		\textbf{-0.83} & ~~~~\textbf{15} & \textbf{h}$_{\rm{max}}$         &0.179 &0.178 &0.177 &0.175 &0.173 &0.170 &0.168 &0.165 &0.162 &0.158 &0.155 \\
		&                 & \textbf{M}$_{\rm{FL}}$ 	     &3.321 &3.309 &3.276 &3.228 &3.170 &3.105 &3.035 &2.962 &2.887 &2.812 &2.737 \\
		&                 & $\boldsymbol{\chi}_{\rm{f}}$   &0.417 &0.414 &0.408 &0.398 &0.386 &0.373 &0.358 &0.343 &0.327 &0.310 &0.294 \\ \cline{2-14}
		& ~~~~\textbf{30} & \textbf{h}$_{\rm{max}}$         &0.359 &0.358 &0.355 &0.351 &0.346 &0.341 &0.335 &0.329 &0.323 &0.317 &0.311 \\
		&                 & \textbf{M}$_{\rm{FL}}$ 	     &3.321 &3.309 &3.276 &3.228 &3.170 &3.105 &3.035 &2.962 &2.887 &2.812 &2.737 \\
		&                 & $\boldsymbol{\chi}_{\rm{f}}$   &0.417 &0.414 &0.408 &0.398 &0.386 &0.373 &0.358 &0.343 &0.327 &0.310 &0.294 \\ \cline{2-14}
		& ~~~~\textbf{60} & \textbf{h}$_{\rm{max}}$         &0.718 &0.716 &0.711 &0.703 &0.693 &0.683 &0.671 &0.659 &0.646 &0.634 &0.621 \\
		&                 & \textbf{M}$_{\rm{FL}}$          &3.321 &3.309 &3.276 &3.228 &3.170 &3.105 &3.035 &2.962 &2.887 &2.812 &2.737 \\
		&                 & $\boldsymbol{\chi}_{\rm{f}}$   &0.417 &0.414 &0.408 &0.398 &0.386 &0.373 &0.358 &0.343 &0.327 &0.310 &0.294 \\ \cline{2-14}
		& ~~~~\textbf{90} & \textbf{h}$_{\rm{max}}$         &1.077 &1.074 &1.066 &1.054 &1.040 &1.024 &1.006 &0.988 &0.970 &0.951 &0.932 \\
		&                 & \textbf{M}$_{\rm{FL}}$ 	     &3.321 &3.309 &3.276 &3.228 &3.170 &3.105 &3.035 &2.962 &2.887 &2.812 &2.737 \\
		&                 & $\boldsymbol{\chi}_{\rm{f}}$   &0.417 &0.414 &0.408 &0.398 &0.386 &0.373 &0.358 &0.343 &0.327 &0.310 &0.294 \\ 
		\hline	
	\end{tabular}\label{table:BN}
\end{table*}

\bsp	
\label{lastpage}
\end{document}